\documentclass{article}
\usepackage{graphicx} 
\usepackage{color}

\usepackage[english]{babel}
\usepackage{amsfonts} 
\usepackage{xparse}
\usepackage{comment}

\usepackage[normalem]{ulem}

\usepackage{diagbox} 
\usepackage[utf8]{inputenc}
\usepackage{makecell} 

\usepackage[colorinlistoftodos,prependcaption,textsize=footnotesize]{todonotes}
\usepackage{xcolor}

\usepackage[shortlabels]{enumitem}
\usepackage{bm}

\usepackage{authblk}
\usepackage[letterpaper,top=2cm,bottom=2cm,left=2.5cm,right=4.5cm,marginparwidth=3.5cm]{geometry}
\usepackage{subcaption}

\usepackage{amsmath}
\usepackage{amsthm}
\usepackage{graphicx}
\newtheorem{theorem}{Theorem}[section]

\newtheorem{remark}[theorem]{Remark}

\usepackage{soul} 

\usepackage{amssymb}

\usepackage[colorlinks=true, allcolors=blue]{hyperref}
\newcommand{\norm}[1]{\left\lVert #1 \right \rVert}

\title{Clustering in co-evolving opinion dynamics: reduced SPDE models}

\author[a,b]{Sebastian Zimper}
\author[a]{Nata\v sa Djurdjevac Conrad}
\author[c]{Federico Cornalba}
\author[d]{Ana Djurdjevac}

\affil[a]{Zuse Institute Berlin, Germany}
\affil[b]{Institut f\"ur Mathematik und Informatik, Freie Universit\"at Berlin, Berlin, Germany}
\affil[c]{Department of Mathematical Sciences, University of Bath, UK}
\affil[d]{University of Oxford, Mathematical Institute, UK}

\begin{document}

\maketitle

\begin{abstract} 
    Clustering is a fundamental collective phenomenon in agent-based models (ABMs) of opinion dynamics. To study clustering in systems with co-evolving social and opinion variables, we derive stochastic partial differential equation (SPDE) models that describe the evolution of clusters. Specifically, these SPDE models are reduced in the sense that their domain accounts only for social variables, while opinion variables are instead incorporated in a multiplicative (Eulerian velocity-type) fashion.The model reduction is primarily guided by computational efficiency, as well as  model interpretability.
    We consider two settings: one in which opinions do not affect social interactions, and another one in which a feedback mechanism couples the two. Our approach extends reduced PDE modelling to a stochastic framework, which is essential for capturing long-term cluster behaviour. Numerical experiments indeed demonstrate that the proposed reduced SPDEs substantially decrease computational cost compared to full-state SPDE models, such as the Dean--Kawasaki equation, while still accurately reproducing the clustering behaviour of the underlying ABM. As a result, these reduced models provide an efficient tool for studying systems with large populations, including those arising in the analysis of real-world data: in particular, we provide an application related to the large-scale General Social Survey (GSS), which comprises opinion and social data of the US population since 1972.
\end{abstract}


\section{Introduction}

Opinion dynamics plays a central role in understanding how beliefs propagate, why societies polarize and which mechanisms lead to the formation of consensus. With the rise of digital social media and increasing availability of large-scale social data, these questions can now be studied quantitatively, by analysing opinion distributions and their temporal evolution on a population level. In particular, using agent-based models (ABMs) we can simulate how individual, microscopic interactions give rise to collective macroscopic system behaviour. In these models, agents represent individuals with a given set of rules that govern their actions and interactions with other individuals and with their environment. Through these interactions the individuals change their opinions and the population exhibits collective phenomena such as consensus, polarization, or fragmentation, all of which represent different types of clustering of like-minded agents. Consensus corresponds to the formation of a single cluster in which all agents share a similar opinion; polarization to a small number of clusters with opposing opinions; and fragmentation to many small clusters in which agents are like-minded. 

Despite their wide use and modelling flexibility, many existing opinion dynamics ABMs face two key limitations: 
\begin{enumerate}
\item \emph{Modelling and empirical verification}: traditional ABMs usually consider over-simplified social interactions and rarely connect to available empirical data.
\item \emph{Computational complexity}: ABMs' computational costs increase rapidly with population size, limiting their applications in real-world scenarios.
\end{enumerate}

Addressing the first limitation, i.e. the gap between theoretical modelling and empirical verification, early ABMs of opinion dynamics, such as those by Hegselmann and Krause \cite{hegselmann2002} and Deffuant et al. \cite{deffuant2000}, assume a well-mixed population where every agent can interact with every other agent. To capture structural constraints of real-world interactions, complex networks have been introduced in opinion dynamics models, linking interaction patterns to underlying social network topology \cite{peralta2022opinion,gabbay2007,SvenEckehard2019}. Different types of complex network representations have been considered, ranging from static to temporal networks, and from synthetic model networks to those inferred from empirical data. Of particular relevance are models of adaptive networks \cite{Kurths2025,kozma2008,ChaosAdaptiveNet17,KurthsEtAl2023Review}, where social connections between agents change based on their opinion similarity, i.e. ties are formed between like-minded agents and ties between agents with opposing views are cut. More general approaches go beyond purely opinion-driven rewiring and consider social dynamics as an additional driving force of the system \cite{djurdjevac2022feedback, djurdjevac2024co}. This enables a more flexible modelling framework for studying how the co-evolution of opinion and social dynamics influence interaction patterns and collective opinion outcomes. In this work, we will use such a setting as our base-line model. Further extensions include incorporating the influence of external actors, such as media or online influencers, on the dynamics \cite{Helfmann2023, zimper2025mean}.

Concerning the second limitation of computational complexity, ABM systems used in applications (or more generally particle systems), typically consist of a large number of agents. As a result, tracking the full state of all agents 
becomes computationally costly. To address this fundamental issue, continuum models describing the evolution of a limited number of aggregate features of the system (such as opinion distributions) are usually deployed as approximate -- but usually cheaper -- model surrogates. Much of the existing literature on surrogate modelling focuses on deriving \emph{mean-field limit} (MFL) type of models \cite{Sznitman1991} that describe the `average' quantities of interest and are typically formulated as partial differential equations (PDEs) whose domain is the agents' variable space. By construction, these models do not capture the effect of agents' fluctuations. However, most long-term collective phenomena are deeply influenced by these fluctuations on the microscopic level (see e.g. the discussion in \cite{Garnier2016}). This has motivated the development of continuum models incorporating noise given in terms of \emph{stochastic} partial differential equations (SPDEs), such as the Dean--Kawasaki equation \cite{Dean1996,Kawasaki1998}. Nevertheless, both PDE and SPDE models remain computationally efficient only in low-dimensional state space (typically for dimensions less than four). To mitigate this limitation, one may further reduce the model by selecting a few of the most important variables and derive the PDE/SPDE surrogate models only for those quantities. We refer to such models as \emph{reduced} PDE/SPDEs, in contrast to \emph{full} models, which describe the evolution of the complete set of variables. The goal  is that these reduced models retain the ability to capture the collective behaviour of the underlying ABMs, while achieving a substantially lower computational cost than the full counterparts. As alluded to, one of the key macroscopic features of interest in opinion dynamics models is the formation and evolution of opinion clusters. An accurate description of clustering phenomena via PDEs/SPDEs is relevant not only to opinion dynamics, but also to a broader range of applications, as highlighted in the following literature review. 

This paper presents several innovations in the mathematical description of clustering phenomena for agent (particle) systems of \emph{opinion dynamics}. We propose novel \emph{reduced} SPDE models for opinion dynamics, and address the two aforementioned limitations of the opinion dynamics in the following ways:
\begin{enumerate}
\item Addressing the gap between modelling and real-world application: In order to account for more realistic dynamics, the baseline ABM, for which we derive reduced SPDE models, includes both opinion and social dynamics. Moreover, it is formulated to account for two relevant cases, without and with feedback loop between the two processes. In contrast to most of the existing literature, both of these cases feature a detailed description of social interaction, and of the interplay between social state and opinion variable. While we study only these two specific types of dynamical coupling, these cases are chosen to provide great flexibility with respect to general formulation of ABM models (see \cite{djurdjevac2022feedback} for a discussion). Furthermore, we deploy our new models on real-world, empirical data from the General Social Survey (GSS) \cite{davern2024gss} and show that these can capture important emerging phenomena. 
\item Addressing the computational complexity: Our proposed SPDE model reduces the computational complexity of the ABM and associated full description, while accounting for the fluctuations necessary for the evolution of clusters. To the best of our knowledge, this is the first derivation of a reduced SPDE model for co-evolving social and opinion dynamics for which the full SPDE was derived in \cite{djurdjevac2022feedback}. Our numerical results demonstrate that using the reduced SPDEs allow for a description of the clustering phenomena at a greatly reduced computational cost.
\end{enumerate}

\subsection*{Literature review}

As the primary innovations of this paper are on the reduced modelling of ABMs, we provide an  overview of the literature in order to contextualize our work. To provide a broader perspective, we do not restrict our attention to clustering phenomena in opinion dynamics, but consider particle systems more generally. Given the focus of this paper, we concentrate on PDE and SPDE approaches, grouping the existing contributions into the categories below. Before delving into these categories, we recall that a full model refers to a continuum description that characterizes the agent density over the entire state space, whereas reduced models describe densities restricted to a lower-dimensional subset of that space.

\begin{itemize}

\item \textbf{PDE/SPDE models of full type}. Mathematical descriptions which use \emph{full} PDEs and SPDEs to describe particle systems have flourished in recent years. Many of these innovations have come in the field of fluctuating hydrodynamics \cite{spohn2012large}, and have contributed both PDE-based results as well as several SPDE-based results. Results on the PDE side include the GENERIC formalism for Vlasov--Fokker--Planck equations \cite{duong2013generic}, propagation of chaos results for mean field kinetic particles \cite{monmarche2017long}, and properties of mean field limits for non-Markovian interacting particles \cite{DuongPavliotis2018}. On the SPDE side, advances include well-posedness and fluctuation properties of conservative SPDEs of fluctuating hydrodynamics \cite{fehrman2019well, dirr2020conservative,Fehrman2023},  high-order quantitative fluctuation bounds for numerical approximations of diffusive/weakly interacting particles \cite{Cornalba2023, cornalba2023density}, quantitative fluctuation bounds and non-negativity for analytical approximations of diffusive/weakly interacting particles  \cite{djurdjevac2024weak,Djurdjevac2026weakerr}, as well as variance reduction methods \cite{cornalba2025multilevel} and hybrid SPDE-particle based techniques \cite{djurdjevac2025hybrid}. There are several works which specifically address clustering phenomena. A linear stability analysis of the mean-field PDE is used in \cite{Wang2017,Garnier2016} to characterize the emergence of clusters. Several contributions employ SPDE frameworks to describe the evolution of clusters in weakly interacting particle systems, including \cite{Wehlitz2025,wehlitz2026data}. Other related studies address clustering in diffusive systems \cite{leimkuhler2025cluster}, feedback effects in opinion dynamics \cite{djurdjevac2022feedback}, and, more recently, provide quantitative analyses of phase transitions, cluster mass exchange, and metastability in MFLs for one-dimensional weakly interacting particles \cite{gerber2025formation}.

\item \textbf{Reduced PDE models}.
Reduced PDE models are typically obtained from the full PDEs through an appropriate closure approximation, often implemented using a \emph{mono-kinetic} ansatz. Examples of this approach include applications to swarming \cite{CarrilloEtAl2010B}, flocking dynamics \cite{Choi2017}, and coupled opinion--preference models without feedback \cite{Pareschi2019}. The latter is particularly relevant to our work, as it captures the formation of clusters in opinion dynamics, although it does not account for stochastic fluctuations. As an example of a reduced PDE model constructed independently of the MFL, we note the deterministic Cucker--Smale PDE studied in \cite{zimper2025reduced}.

\item \textbf{Reduced SPDE models}.
Reduced SPDE models have been extensively used in the context of fluctuating hydrodynamics for underdamped inertial systems, see applications in chemistry \cite{lutsko2012dynamical}, biology \cite{thompson2011lattice}, and active matter \cite{cates2015motility}. Aspects concerning the rigorous mathematical analysis of such models have been addressed in recent years: in particular, we mention the introduction of the \emph{Regularised Inertial Dean--Kawasaki} (RIDK) model for Langevin dynamics in particles' position and velocity \cite{Cornalba2019,cornalba2020weakly,cornalba2021well, cornalba2023regularised}, as well as a recent application of the RIDK model to computational chemistry \cite{jin2025field}. In contrast to the cited works, our ABM has noise in both components and there is no obvious way to interpret the opinion as a  velocity variable. 

\end{itemize}
While they are not directly linked to the methodologies of this paper, we refer to data-driven approaches for opinion dynamics in \cite{Albi2024data} (quantifying the effect of social media networks on opinion dynamics phenomena such as polarization, and emergence of collective behaviour using evolutionary and kinetic models), and characterization of clustering behaviour via data-driven reduction of transfer operators \cite{wehlitz2026data}.

\subsection*{Outline}

The rest of the paper is organised as follows. In Section \ref{sec_particle_systems} we present the underlying ABM,  for which in Section \ref{sec:ReducedSPDE} we introduce our model reduction approach and derive reduced SPDE models. Section \ref{sec:NumericalResults} contains numerical results demonstrating the ability of the reduced SPDE models to mimic the clustering behaviour of the underlying ABM as well as a comparison to real-world data. Lastly our conclusions and outlook are presented in Section \ref{sec:Conclusion}.

\section{ABM formulation of opinion and social dynamics}\label{sec_particle_systems}

We study a closed system of $N \in \mathbb{N}$ interacting agents with a focus on interplay between social and opinion dynamics, as introduced in \cite{djurdjevac2022feedback}. At a time $t \geq 0$ the state of the $i$-th, $i = 1, \dots, N$, agent is described by two variables:
\begin{itemize}
    \item the `social space' variable $X^i_t \in \mathbb{T}^d$, $d \geq 1$, which identifies the agent's `position' in a social space, determined by, e.g., socio-demographic attributes and ideological stances. We assume that the social space is given by the $d$-dimensional unit torus to simplify some of the mathematical derivation (i.e., manipulations relating to integration by parts). The generalization to other spaces is left for future work. 
    \item 
    `opinion space' variable $\theta^i_t \in \mathbb{R}$ which quantifies the opinion of an agent for a specific topic, e.g. attitudes on governmental issues, environmental risk perception, policy support. Positive values of this variable indicate agreement with the topic, and negative values disagreement. Since we consider only one topic, the opinion space is one-dimensional. Our approach can however be extended to a multi-dimensional opinion space, see minimal discussion in Appendix \ref{sec:App_NoFeedback}.
\end{itemize}

We denote the collection of all social and opinion space variables at $t$ by $\textbf{X}_t := (X^i_t)_{i=1}^{N}$ and $\bm{\Theta}_t := (\theta^i_t)_{i=1}^{N}$, respectively. Following the notation introduced in \cite[Eqn.~(1)]{djurdjevac2022feedback}, a fairly general model encompassing social and opinion dynamics considered in this work, takes the form 
\begin{align}
    \label{eq_GeneralParticle}\tag{Op-Dyn}
    \begin{aligned}
    d\textbf{X}_t    & =  \tilde{U}(\textbf{X}_t,\bm{\Theta}_t) dt + \tilde{\sigma}_{s}(\textbf{X}_t,\bm{\Theta}_t) d\mathbf{W}_t ,\\
    d\bm{\Theta}_t    & =  \tilde{V}(\textbf{X}_t,\bm{\Theta}_t) dt + \tilde{\sigma}_{o}(\mathbf{X}_t,\bm{\Theta}_t) d\mathbf{B}_t  ,
    \end{aligned}
\end{align}
for suitable functions $\tilde{U}$ and $\tilde{V}$ describing how agents' positions and opinions influence their evolution in the social and opinion spaces, together with diffusion coefficients $\tilde{\sigma}_s$ and $\tilde{\sigma}_o$, and independent Brownian motions $\mathbf{W}=(W^i_t)_{i=1}^{N}$ and $\mathbf{B}=(B^i_t)_{i=1}^{N}$. The interaction functions ad diffusion coefficients could also have an explicit time-dependence. This model enriches typical opinion dynamics by considering the impact of social ties, such that agents' opinions evolve under the influence of both social dynamics and opinions of other agents. Concerning the social dynamics, in this work we focus on two specific systems: 
\begin{itemize}
    \item in the first system, social dynamics evolves independently of the opinion dynamics. This type of dynamics is present in systems where social ties are given by fixed organizational structure or physical proximity of agents, i.e. independent of the agents' opinions, see Section \ref{subsub_no_f} for the model formulation; 
    \item in the second system, a feedback loop between social and opinion dynamics is present. Such dynamics arises in online social networks or political communities where opinions influence social interactions, see Section \ref{subsub_f} for more details.
\end{itemize}
These systems provide a broad framework for studying cluster formation in the social space and the corresponding opinion patterns within the resulting clusters.

\subsection{First system: non-feedback model}\label{subsub_no_f}

The first model we consider describes how the formation of clusters in the social space shape the opinions of their members. In this setting, the social dynamics affect the opinion dynamics, while the opinion dynamics does not impact the social interactions. We will refer to this model as the \emph{non-feedback} model, since the positions in the social space influence the opinions but not vice-versa. This system reads
\begin{equation}
    \label{particles_no_feedback}\tag{NF}
    \begin{aligned}
        dX^i_t &= - \frac{1}{N} \sum_{j=1}^N b(X^i_t-X^j_t)dt - \nabla V(X_t^i) dt + \sigma_s dW_t^i, \\
        d\theta^i_t &= - \frac{1}{N} \sum_{j=1}^N a(X^i_t-X^j_t) (\theta^j_t-\theta^i_t) dt + \sigma_o dB_t^i,
    \end{aligned}
\end{equation}
where $b: \mathbb{T}^d \to \mathbb{R}^d$ and $a:  \mathbb{T}^d \to \mathbb{R}$ are pairwise interaction potentials between agents, $V:  \mathbb{T}^d \to \mathbb{R}$ is an environment potential in the social space, and the noise is given by independent $\mathbb{R}^d$-valued Brownian motions $(W^i)_{i=1}^{N}$ and $\mathbb{R}$-valued Brownian motions $(B^i)_{i=1}^{N}$, with strengths determined by the non-negative constants $\sigma_s$ and $\sigma_o$. One could also consider the setting where the noise enters in the multiplicative way, as introduced in \cite{djurdjevac2022feedback}. Note that in \eqref{particles_no_feedback}, the interaction maps $\tilde{U}$ and $\tilde{V}$ from \eqref{eq_GeneralParticle} are given as averaged sums of the pairwise interactions $b$ and $a$ over all agents. This formulation renders the agents indistinguishable as they influence the dynamics only through  aggregated quantities. Typically $b$ and $a$ are chosen such that agents who are more similar experience a stronger mutual attraction. This models the sociological concept of homophily, namely that people prefer others which are similar to themselves. The potential $V$ does not depend on the positions of the population, but serves to describe the `shape' of the social space, encoding which social stances are currently popular.

\subsection{Second system: feedback model}\label{subsub_f}

The second model describes not only how social dynamics shape opinions, but also how opinions influence the interactions in the social space. This provides a more realistic representation of the intricate interplay between individuals' views and their patterns of communication. The system, which we refer to as the \emph{feedback} model, reads
\begin{equation}\label{particles_feedback}\tag{F}
    \begin{aligned}
        dX^i_t &= -\frac{1}{N} \sum_{j=1}^N b(X^i_t-X^j_t) \text{sgn}(\theta^i_t \theta^j_t) dt - \nabla V(X_t^i) dt + \sigma_s dW_t^i, \\
        d\theta^i_t &= - \frac{1}{N} \sum_{j=1}^N a(X^i_t-X^j_t)  (\theta^j_t-\theta^i_t) dt + \sigma_o dB_t^i .
    \end{aligned}
\end{equation}
This model also includes a non-linear term $\text{sgn}(\theta^i_t \theta^j_t)$ that incorporates attraction/repulsion for any agent pair $i,j$ based on whether their opinions are of the same stance, i.e. whether $\theta^i_t,\theta^j_t$ have the same sign. This feature -- in contrast to \eqref{particles_no_feedback} -- means that opinions directly influence the agents' positions in the social space, creating a full feedback loop between opinion and social dynamics. For a multi-dimensional opinion space, the feedback can be generalised to $\text{sign}(\theta^i_t \cdot \theta^j_t)$.

\section{Reduced SPDEs}
\label{sec:ReducedSPDE}

\subsection{Towards a reduced SPDE model of the ABM}

For both ABMs introduced in the previous section, sufficiently strong interactions (or, equivalently, low diffusion) lead to the formation of clusters in the social space. In particular, it is assumed throughout that the system parameters lie in a regime where clustering occurs in the ABM. The long-time behaviour of these clusters can be described by the Dean--Kawasaki-type equations, which provide a full SPDE representation of the particle systems' fluctuations \cite{djurdjevac2022feedback}. To further reduce the computational cost, we derive reduced SPDE models. This work builds upon -- and adds significant modelling innovations to -- the reduced models so far developed by the authors for fluctuating hydrodynamics \cite{cornalba2021well} and deterministic flocking dynamics \cite{zimper2025reduced}. Here, we only provide a high-level description of our the nature of our approach, before spelling out the derivation of our reduced SPDE models for the ABMs \eqref{particles_no_feedback} and \eqref{particles_feedback} in Appendix \ref{sec:App_feedback}--\ref{sec:App_NoFeedback}.

We perform model reduction at the level of the opinion information $\bm{\Theta}$, such that our proposed reduced SPDE models give a continuum description of the social positions $\mathbf{X}$. This provides us with a reduced description of how agents cluster in the social space as well as what opinions are held within those clusters. Model reduction serves the primary purpose of drastically reducing the computational cost with respect to full models. 
We achieve model reduction by considering suitable continuous densities which are defined only on the social-space: The baseline densities are the empirical density $\rho$ and opinion-weighted empirical density $j$ given by 
\begin{align}\label{densities}
\rho(x,t) := N^{-1}\sum_{i=1}^{N}{\delta(x-X^i_t}),\qquad j(x,t) := N^{-1}\sum_{i=1}^{N}{\theta^i_t\delta(x-X^i_t)}.
\end{align} 


The former characterizes the density of agents in the social space, while the latter is the corresponding opinion-weighted density, analogous to a momentum or flux variable in continuum models. By their very definition in \eqref{densities}, the densities enjoy a far tighter and more direct connection to the agent dynamics than the one which arises in more classical approaches where reduction is achieved from the full MFL dynamics, rather than directly from the ABM. Model reductions from full models \cite{CarrilloEtAl2010B,Choi2017} typically heavily rely on the already mentioned \emph{mono-kinetic} ansatz,  which if often considered on a mean-field level. However, we deploy it for our finite size particle system formulation. Specifically, we postulate that the joint (full) density $f(x,\theta,t) = N^{-1}\sum_{i=1}^{N}{\delta(x-X^i_t)\delta(\theta - \theta^i_t)}$ is directly aligned to the limiting marginal $\rho(x,t)$ introduced in \eqref{densities} by means of the relation $f(x,\theta,t)=\rho(x,t)\delta(\theta - u(x,t))$, where $u(x,t) = j(x,t)/\rho(x,t)$. This ansatz is heuristically satisfied when agents which are close in the social space share similar opinions. Performing such a reduction from the deterministic model obtained from the MFL, means that the fluctuations of the ABM cannot be reproduced in the reduced model. Although we do use the mono-kinetic ansatz in our approach for the feedback model \eqref{particles_feedback} this is done directly for the ABM, such that the fluctuations are still accounted for.

The advantages and flexibility of our model reduction (which is hinged on \eqref{densities}, and which, as such, bypasses the MFL) have been studied to a relevant -- but still rather limited extent -- in specific applications of fluctuating hydrodynamics \cite{Cornalba2019,cornalba2021well,cornalba2023regularised,jin2025field} and a test-case scenario in deterministic flocking dynamics \cite{zimper2025reduced}. However, prior to this work, such a reduction strategy had not been applied to realistic (and mathematically challenging) ABMs in opinion dynamics, such as \eqref{particles_no_feedback} and \eqref{particles_feedback}. 

While precise details of the reduced SPDE models associated with ABMs \eqref{particles_no_feedback} and \eqref{particles_feedback} will follow (see \eqref{spde_ReducedNoFeedSPDE} and \eqref{spde_ReducedFeedbackSPDE}), for now it suffices to say that these SPDE models take the following abstract form
\begin{equation}
    \label{eq_GeneralSPDE}\tag{Red-SPDE}
    \begin{aligned}
    \partial_t \rho    & =  \text{drift}_\rho + N^{-\frac{1}{2}}\text{noise}_{\textbf{X},\text{div}}  \\
    \partial_t j & =   \text{drift}_j +  N^{-\frac{1}{2}} \text{noise}_{\bm{\Theta}} +  N^{-\frac{1}{2}} \text{noise}_{\textbf{X},\text{div}} \\
    \end{aligned}
\end{equation}
where 
\begin{equation}
    \label{eq_GeneralDrift}\tag{Drift}
    \begin{aligned}
    \text{drift}_\rho   & =  \text{diffusion}_\rho + \text{convolution}_\rho  \\
    \text{drift}_j   & =  \text{diffusion}_j + \text{convolution}_j .
    \end{aligned}
\end{equation}
As the above abbreviations suggest, the SPDE models considered here have the following structural features: (i) all deterministic features, i.e. the drift terms, arise  either from  diffusion or from averaging relevant fields via convolutions against suitable potentials; (ii) noises may be driven by either the state-space or opinion agent dynamics; (iii) in addition, certain noise terms may be mass-preserving (through a divergence operator).

As we will see, $\rho$ and $j$ are  not  sufficient to fully characterize the systems dynamics,  and we will therefore introduce additional continuous densities into \eqref{eq_GeneralSPDE} when needed. These additional densities will nonetheless adhere to the reduction ansatz and depend only on the social-space variable.

\begin{remark}\label{remark-double-notation}
In the interest of not burdening notation, the symbol we use for any given empirical measure is reused verbatim for its counterpart in relevant stochastic equations. For instance, look at the use of $\rho,j$ (used as empirical density in \eqref{densities}, and also as unknowns in the stochastic equation \eqref{eq_GeneralSPDE}).
\end{remark}

\subsection{Derivation of the reduced SPDE: Non-feedback model}

For the non-feedback model \eqref{particles_no_feedback}, the reduced model we study is the following stochastic PDE of Dean--Kawasaki type (see \cite{Dean1996, Kawasaki1998})
\begin{equation}
    \label{spde_ReducedNoFeedSPDE}\tag{SPDE-NF}
    \begin{aligned}
        \partial_t \rho &= \frac{\sigma_s^2}{2} \Delta \rho + \nabla \cdot [\rho (b *\rho + \nabla V)] + N^{-\frac{1}{2}} \sigma_s \nabla \cdot (\sqrt{\rho} \xi_\rho), \\
        \partial_t j &= \frac{\sigma_s^2}{2} \Delta j - \rho (a *j) + j (a * \rho) + \nabla \cdot [j (b *\rho + \nabla V)] + N^{-\frac{1}{2}} \sigma_o \sqrt{\rho} \xi_j + N^{-\frac{1}{2}}  \sigma_s \nabla \cdot (\sqrt{K} \xi_\rho), \\
        \partial_t K &= \frac{\sigma_s^2}{2} \Delta K - 2 j (a*j) + 2 K (a * \rho) + \sigma_o^2 \rho  + \nabla \cdot [K (b *\rho + \nabla V)] ,
    \end{aligned}
\end{equation}
where $\xi_\rho$ and $\xi_j$ are independent vector and scalar space-time white noises, respectively. We defer the full justification of \eqref{spde_ReducedNoFeedSPDE} to Appendix \ref{sec:App_NoFeedback}. We note that the time-evolution of $\rho$ is governed by the standard Dean--Kawasaki equation for interacting particle systems, which is derived by an application of It\^o's formula, properties of the convolution, and by matching the correlation of the noise terms. The reduced model \eqref{spde_ReducedNoFeedSPDE} is posed on a $d$ dimensional space, as opposed to $d+1$ dimensional space for full models.
We refer to \cite{Dean1996} for a detailed derivation and discussion. A similar computation gives the equation for $j$, where the final term 
\begin{equation}
    K(x,t) := \frac{1}{N} \sum_{i=1}^N (\theta_t^i)^2 \delta(x - X_t^i) 
\end{equation}
is, however, not closed in terms of  $\rho$ and $j$. In the non-feedback case, a closing approximation is therefore only required for the noise terms. This closure is achieved by introducing a hierarchy of densities only dependent on social-space variables as follows. The time-differential of $K$ is computed as
\begin{equation*}
    \begin{aligned}
    \partial_t K &= \frac{\sigma_s^2}{2} \Delta K - 2 j (a*j) + 2 K (a * \rho) + \sigma_o^2 \rho  + \nabla \cdot [K (b *\rho + \nabla V)] \\
    & + 2 \sigma_o N^{-\frac{1}{2}} \sqrt{K} \, \xi_K + N^{-\frac{1}{2}} \sigma_s \nabla \cdot (\sqrt{L} \, \eta_K) ,
    \end{aligned}
\end{equation*}
where $\xi_K$ and $\eta_K$ are space-time white noises, and
\begin{equation}
    L := \frac{1}{N} \sum_{i=1}^N (\theta_t^i)^4 \delta_{X_t^i} .
\end{equation}
It is clear that if we continue with this scheme (i.e., if we derive an equation for $L$ as well) we will obtain noise terms containing higher powers of $\theta_t^i$, and which cannot be closed with the existing set of densities. The crucial point here is that such hierarchy of approximations is conducted -- exclusively -- at the noise level, and is thus stable on the account of the SPDE noises enjoying small prefactors of $N^{-1/2}$ given by averaging over the number of agents: 
 in particular, these prefactors of $N^{-1/2}$ cumulate the further down we go in the hierarchy of equations, and this will help guide which noises should be kept and which ones should be dropped. 
Since we will only use the reduced SPDE for $N \gg 1 $, we drop the noise terms in $K$ thereby obtaining the closed model \eqref{spde_ReducedNoFeedSPDE}. Our approach in this case allows us to bypass mono-kinetic approximations entirely and thereby avoid the density divisions $u(x,t)=j(x,t)/\rho(x,t)$ from it. These divisions are undesirable as they may lead to issues when performing numerical discretisations due to the fluctuations in low densities regimes. A detailed derivation of all terms is given in Appendix \ref{sec:App_NoFeedback}.

By conducting extensive numerical simulations in Section \ref{sec:NumericalResults}, for both moments and order-statistics quantities, we verify that our SPDE is effective in reproducing the macroscopic clustering behaviour of the ABM at a greatly reduced computational cost when compared to the ABM with a large number of agents and the full SPDE description given by the Dean--Kawasaki equation. Furthermore, we observe that our method outperforms noise-linearised SPDE analogues. While our SPDE model currently performs well in reproducing the ABM, it is important to mention that the current noise hierarchical procedure is statistically exact in reproducing the mean and covariance of the particle noise, but not the correlations. In particular, although the quadratic variation of the noise terms in the SPDE matches those of the ABM, it is not evident how to represent them so as to reproduce the correlations between Brownian motions in the particle system. We therefore adopt a specific choice of noise terms in the SPDE, recognizing that this may not fully preserve the correlation structure of the ABM. A more detailed discussion is provided in Appendix~\ref{sec:App_NoFeedback}. While this is a preliminary modelling choice that we have taken to produce a minimally working example \eqref{spde_ReducedNoFeedSPDE}, such an example is already performing very well.

\subsection{Derivation of the reduced SPDE: Feedback model}
\label{subsec:FeedbackSPDE}

The reduced SPDE description for the feedback ABM \eqref{particles_feedback} is given by
\begin{equation}
    \label{spde_ReducedFeedbackSPDE}\tag{SPDE-F}
    \begin{aligned}
        \partial_t \rho =& \frac{\sigma_s^2}{2} \Delta \rho + \nabla \cdot \left[ \rho  ( b * ( \rho \, \text{sgn}(j) ) \text{sgn}(j) + \nabla V ) \right] + N^{-\frac{1}{2}} \sigma_s \nabla \cdot (\sqrt{\rho} \xi_\rho) \\
        \partial_t j =& \frac{\sigma_s^2}{2} \Delta j - \rho (a *j) + j (a * \rho) + \nabla \cdot \left[  j  ( b * ( \rho \, \text{sgn}(j) ) \text{sgn}(j) + \nabla V  )  \right] \\
        &+ N^{-\frac{1}{2}} \sigma_o \sqrt{\rho} \xi_j + N^{-\frac{1}{2}}  \sigma_s \nabla \cdot  (u\sqrt{\rho} \xi_\rho) \\
        u =& j / \rho ,
    \end{aligned}
\end{equation}
where $\xi_\rho$ and $\xi_j$ are vector and scalar space-time white noise, respectively. Obtaining this reduced description is significantly more complicated than the previous case due to the sharp nonlinearity terms $\text{sgn}(\theta^i_t \theta^j_t)$. These terms mean that not only the noise terms, as in the non-feedback setting, but also the drift terms need to be closed. To illustrate this issue, we provide a brief sketch for the one-dimensional case ($d = 1$). A formal application of It\^o's formula to $\rho$ and $j$, as defined in \eqref{densities}, yields
\begin{equation*}
    \begin{aligned}
        d \rho(x,t)  = & \frac{\sigma_s^2}{2} N^{-1} \sum_{i=1}^N \Delta \delta(x-X_t^i)dt - N^{-1} \sum_{i=1}^N \nabla \delta(x-X_t^i) \cdot dX_t^i \\
        d j(x,t)  = & \frac{\sigma_s^2}{2} N^{-1} \sum_{i=1}^N \theta_t^i \Delta \delta(x-X_t^i)dt - N^{-1} \sum_{i=1}^N \theta_t^i \nabla \delta(x-X_t^i) \cdot dX_t^i \\
        & + N^{-1} \sum_{i=1}^N\delta(x-X_t^i) d\theta_t^i .
    \end{aligned}
\end{equation*}
When inserting $dX_t^i$ from \eqref{particles_feedback} into the above expression it is clear that we will obtain two drift terms with the non-linearity, namely
\begin{equation*}
    \begin{aligned}
        T_1 &:=   \frac{1}{N^2} \sum_{i=1}^N  \nabla \delta(x-X_t^i) \cdot\sum_{j=1}^N  b(X_t^i-X_t^j) \text{sgn}(\theta_t^i \theta_t^j) dt , \\
        T_2 &:=  \frac{1}{N^2} \sum_{i=1}^N \theta^i \nabla \delta(x-X_t^i) \cdot  \sum_{j=1}^N  b(X_t^i-X_t^j) \text{sgn}(\theta_t^i \theta_t^j) dt.
    \end{aligned}
\end{equation*} 
Since terms containing $\text{sgn}(\theta_t^i \theta_t^j)$ cannot be expressed in terms of $\rho$ and $j$, the two above expressions cannot be closed.

To derive the associated SPDE \eqref{particles_feedback}, we perform a mono-kinetic approximation directly on the reduced quantities \eqref{densities}, not on the deterministic MFL dynamics. That is, for the full empirical measure of the system,
\begin{equation}
    \label{eq:FullEmpMeasure}
    f(x,\theta,t) := N^{-1} \sum_{i=1}^N \delta(x-X^i_t) \delta(\theta - \theta_t^i) ,
\end{equation}
we will assume that $f$ has the form
\begin{equation}
    \label{eq:MonoApprox}
    f(x,\theta,t) = \rho(x,t) \delta(\theta - u(x,t)) .
\end{equation}
By the definition of \eqref{eq:FullEmpMeasure}, and using the mono-kinetic approximation \eqref{eq:MonoApprox}, we obtain
\begin{equation*}
    \begin{aligned}
        T_1 &= \int \nabla \delta(x-z) \cdot \int b(z-y) \text{sgn}(\theta' \theta) f(y,\theta,t) dy d\theta f(z,\theta',t) dz \, d\theta' \, dt \\
        & \approx \int \nabla \delta(x-z) \cdot  \int b(z-y) \text{sgn}(u(z) u(y)) \rho(y) dy \, \rho(z) \,  dz \, dt ,
    \end{aligned}
\end{equation*}
and 
\begin{equation*}
        T_2  \approx \int \nabla \delta(x-z) \cdot  \int b(z-y) \text{sgn}(u(z) u(y)) \rho(y) dy \, j(z) \,  dz \, dt .
\end{equation*}
Moreover, assuming that $\rho >0$, we have $\text{sgn}(u(y)\, u(x) ) = \text{sgn}(j(y)) \, \text{sgn} ( j(x) )$. It therefore follows that the terms $T_1$ and $T_2$ can be rewritten as the second term in the equation \eqref{spde_ReducedFeedbackSPDE} for $\rho$ and the fourth term in the equation for $j$, respectively. In this manner the terms $\text{sgn}(\theta_t^i \theta_t^j)$ in $T_1$ and $T_2$ have been replaced with $\text{sgn}(j(X_t^i) j(X_t^j))$. As a result,  the closing error for the drift terms of the SPDE is quantified by 
\begin{equation}
    \label{eq:ClErr}
    C_E(t) :=N^{-2} \sum_{i,k=1}^N \left| \text{sgn}(\theta^i_t \theta_t^k) - \text{sgn}(j(X^i_t) \, j(X_t^k))  \right| ,
\end{equation}
which has a maximum value of $2(1-1/N)$ and a minimum value of zero. The closing error is small when the sign of an agent's opinion is well represented by the sign of the local opinion-weighted density $j$ at its position in social space. In particular, this occurs when agents that are close in the social space tend to have opinions of the same sign. In this case, the signs of the microscopic interaction terms $\theta_t^i\theta_t^k$ are well approximated by those of the corresponding closure terms $j(X_t^i)j(X_t^k)$. Although the closing error may initially be large, the opinion dynamics tend to align the opinions of agents that are close in the social space, and we therefore expect the error to decrease as the system evolves. In the next section, we track \eqref{eq:ClErr} to assess the accuracy of the closure approximation.

Since the closure is performed at the level of the agent dynamics, the resulting reduced model remains stochastic and is therefore able to capture the evolution of clusters in the ABM (within appropriate regimes). Moreover, our approach is not limited to co-evolving social and opinion dynamics and can be extended to other ABMs in which multiple quantities co-evolve. Lastly, let us also note that by using the mono-kinetic ansatz to close the noise terms we are also able to account for the correlations between the SPDE noise terms exactly. For a detailed (formal) derivation see Appendix \ref{sec:App_feedback}.

The derivation is understood at a formal level for several reasons. First, even at the particle level \eqref{particles_feedback}, the drift is rough: the factor $\operatorname{sgn}(\theta_t^i\theta_t^j)$ is discontinuous in the internal variables, so the interaction is not smooth with respect to $(\theta^i,\theta^j)$. This is precisely where the nonlinearity in $\theta^i\theta^j$ causes difficulty: the drift jumps across the set $\{\theta^i\theta^j=0\}$, and hence standard smooth-coefficient SDE theory does not apply directly. In the presence of non-degenerate Brownian noise, one may nevertheless expect a regularizing effect; this goes back to the classical works of Zvonkin \cite{Zvonkin1974} and Veretennikov \cite{Veretennikov1981}, and was later extended to more singular drift classes by Krylov--R\"ockner \cite{Krylov2004}. Second, if the interaction kernels $a$ and/or $b$ are themselves singular, then the system also falls within the broader framework of mean-field limits for Brownian particles with singular interactions; see for instance Duerinckx--Jabin \cite{Duerinckx2025} and the references therein. Third, the Dean-type computation is itself only formal, since it applies It\^o's formula to empirical Dirac masses such as $\delta(x-X_t^i)$, which are distributions rather than $C^2$ functions. Finally, in the sign-dependent model, any attempt to derive closed equations for moments involving $\operatorname{sgn}(\theta)$ or $|\theta|$ introduces additional non-smoothness in the $\theta$-variable, so further closure assumptions are required.

Although the reduced model is more complicated than that introduced in the previous section \eqref{spde_ReducedNoFeedSPDE}, it represents an important first step toward a continuous description of the ABM \eqref{particles_feedback}, which incorporates a realistic attraction/repulsion term based on opinion alignment, as highlighted in \cite{djurdjevac2022feedback,djurdjevac2024co}. At the same time, as mentioned in the previous section, the density divisions $u(x,t)=j(x,t)/\rho(x,t)$  gives rise to  numerical instabilities, which in turn impact  how accurately  the clustering behaviour of the associated ABM can be reproduced. Developing stable numerical schemes for such systems is therefore an important direction for future work.

\section{Numerical results}
\label{sec:NumericalResults}

In this section we study the accuracy with which the reduced continuum models reproduce the macroscopic/collective behaviour of the underlying ABMs. The particular behaviour we are interested in is the emergence and evolution of clusters.
Throughout this section, we adopt the same interaction forces as in \cite{djurdjevac2022feedback, djurdjevac2024co}, i.e.
\begin{equation*}
    \begin{aligned}
        a(x_i - x_j) & = - \alpha \mathbb{I}_{R_o}(\norm{x_i - x_j}), \\
        b(x_i - x_j) & =  \beta (x_i - x_j) \mathbb{I}_{R_s}(\norm{x_i - x_j}) ,
    \end{aligned}
\end{equation*}
where $\mathbb{I}_r(\cdot)$ is the indicator function for a ball of radius $r$ centred at the origin. The parameters $\alpha , \beta >0$ determine the strength of the interaction, while $R_o, R_s > 0$ describe the interaction radii for opinion and social dynamics, respectively. The evolution in the social space is given by a standard bounded-confidence model in which agents are attracted towards each other if they are within the interaction radius $R_s$ of each other. Similarly, in the opinion space agents only adapt their opinions if they are sufficiently close in the social, not opinion, space. Unless otherwise stated, the following parameter values are chosen:
\begin{equation*}
    \alpha = \beta = 10 , \quad R_s = R_o = 0.1 , \quad \sigma_s = \sigma_o = 0.05 .
\end{equation*}
With this choice of parameters the attractive force in the social space is sufficiently high (or equivalently the magnitude of the noise is sufficiently low)  for clusters to form. Moreover, the dynamics in the opinion space operate at a similar time-scale to those of the social space. We consider the state space $(x,\theta) \in \mathbb{T}^d \times \mathbb{R}$ and a total of $N=10^3$ agents.

To assess how accurately the reduced continuum models  reproduce the dynamics of the ABMs, we use numerical simulations to provide a quantitative comparison between the clustering behaviour of the ABMs \eqref{particles_no_feedback} and \eqref{particles_feedback} with that of the corresponding reduced models \eqref{spde_ReducedNoFeedSPDE} and \eqref{spde_ReducedFeedbackSPDE}. Simulations of the ABMs are performed using the Euler--Maruyama scheme, while for the SPDEs the spatial variable is discretised using a finite-difference scheme and the resulting system of coupled SDEs is solved using the Euler--Maruyama scheme, as described in \cite{Cornalba2023}.

To quantify the evolution of the clusters, we consider three quantities: the relative frequency of clusters, the order parameter 
\begin{equation}
    \label{eq:OrderParam}
    Q_C(t) := \int_{\mathbb{T}^d} \int_{\mathbb{T}^d} \rho(x,t) \rho(y,t)  \mathbb{I}_{R_s}(\norm{x - y}) dx dy ,
\end{equation}
and the opinion parameter, 
\begin{equation}
    \label{eq:OpinionParam}
    Q_o(t) := \int_{\mathbb{T}^d} \left( u(x,t) - \bar{\theta}(t)  \right)^2 \rho(x,t) dx ,
\end{equation}
where $\bar{\theta}$ is the average opinion of all agents. These quantities are computed for numerous realisations of the noise driving the system. 

To compute the cluster frequency, clusters must first be identified from either the empirical measure of the ABM or the solution $\rho$ of the SPDE. We adopt the cluster detection method from \cite{Wehlitz2025} which is briefly described. Since clusters correspond to local maxima of $\rho$, the peak detection function \textit{find\_peaks\_cwt} of the Python package \textit{scipy.signal} is used to locate them. However, due to the stochastic fluctuations, this may identify local maxima in which the density is in fact too small to classify as a cluster. To avoid this, only local maxima which are above a threshold value of $1$ are counted as clusters. The order parameter $Q_C$, introduced in \cite{Wang2017} to study the clustering of the Hegselmann--Krause model, measures the degree of clustering in the system. It is maximised at $Q_C = 1$ when all agents are in a single cluster, and takes the value $Q_C = 1/n$ when there are $n$ clusters of equal size. For further discussion, see \cite{goddard2022noisy}. The opinion parameter $Q_o$ measures the dispersion of opinions. It is minimised when the same opinion is held throughout the whole system, and increases if there is a large density of agents in the social space which have opinions far from the average. In this way, it captures how diverse the opinions are within and across clusters.

\subsection{Non-feedback model}
\label{sec:NumRes_NonFeed}

\begin{figure}[tbh]
    \centering
    \includegraphics[width=0.99\linewidth]{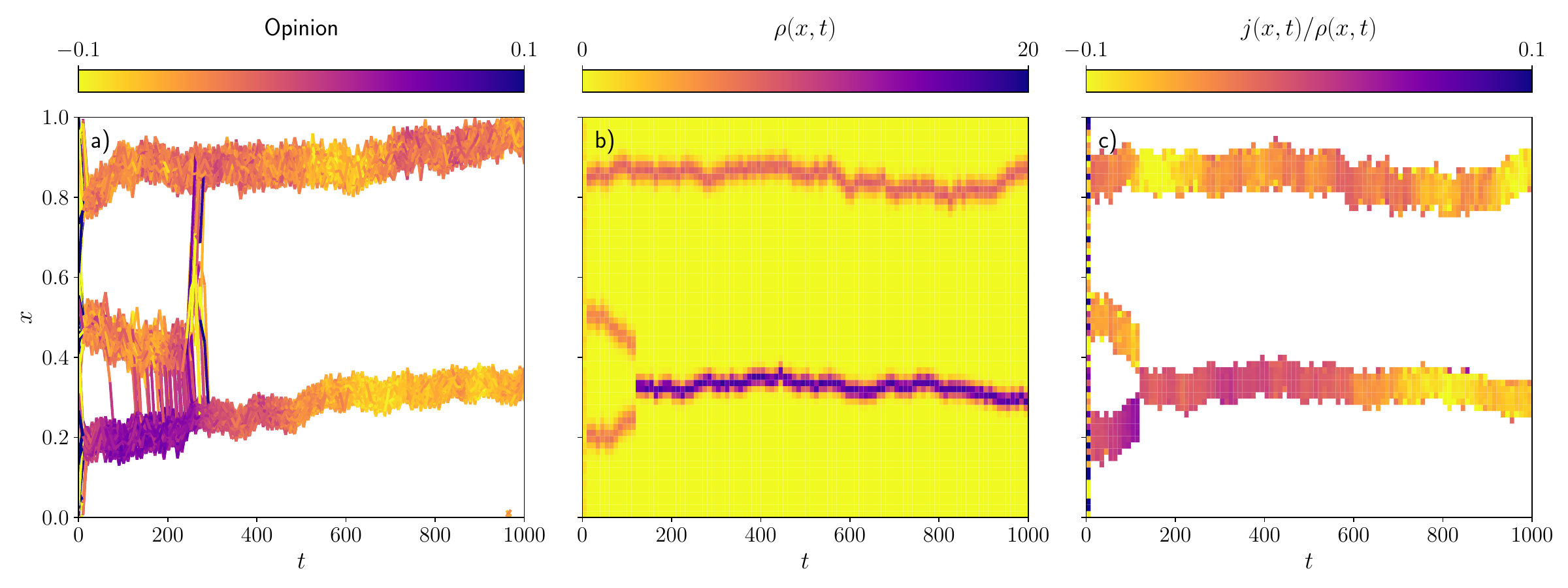}
    \caption{Individual simulations showing the evolution of clusters. The dynamics are shown for $N=1000$ agents on the social space $\mathbb{T}$ and opinion space $\mathbb{R}$ for a total time of $T=1000$. The initial conditions are generated uniformly on $x \in \mathbb{T}$ and $\theta \in [-1,1]$. a) Trajectories of the non-feedback ABM \eqref{particles_no_feedback}. The positions of the agents in the social space are shown, with each agent coloured according to their opinions as given by the colour-bar on the top of the plot. b) The numerical solution $\rho(x,t)$ of the non-feedback \eqref{spde_ReducedNoFeedSPDE}. Regions with more mass are darker as indicated by the colour-bar. c) The local, mean opinion $u(x,t) = j(x,t)/\rho(x,t)$ of the non-feedback \eqref{spde_ReducedNoFeedSPDE}. Note that the colour-bar for $u(x,t)$ is the same as that for the opinion in a)}
    \label{fig:1DsingleSim}
\end{figure}

We begin by showing a simulation of the non-feedback ABM \eqref{particles_no_feedback} and associated reduced SPDE model \eqref{spde_ReducedNoFeedSPDE} with no potential $V$ in Fig.~\ref{fig:1DsingleSim}. The initial data of the ABM is sampled from the uniform distribution on the state space and the SPDE is initialised using the smoothed empirical measure of this. For the ABM, panel a), the positions of the agents in the social space are given for the time interval $t \in [0,1000]$ with each agent coloured according the their opinion. Darker colours correspond to a more positive opinion (e.g. agreement), while brighter colours indicate a negative opinion (e.g.disagreement). The solution of the SPDE, $\rho(x,t)$, and the density ratio $u(x,t) = j(x,t)/\rho(x,t)$ are shown in the panels b) and c), respectively. The same colour-map used for the opinions is used to indicate negative and positive values of $u$. When $\rho(x,t)$ is below a threshold of $0.1$ the corresponding values of $j(x,t)/\rho(x,t)$ are plotted in white, since we are primarily interested in regions containing a substantial density of agents. Although the low-density regions could also be displayed, doing so reduces the visual clarity of the plot. In particular, the normalisation by $\rho$ makes the opinion variations in the clusters difficult to discern. From these simulations we see that qualitatively the ABM and SPDE have similar behaviour. Clusters in the social space are formed very early on in the simulations. The clusters then move throughout the social space and eventually merge. Within the clusters a local consensus is reached which fluctuates due to the noise on the opinion dynamics. When two clusters merge, the opinions in the resulting cluster form a new local consensus which then continuous to fluctuate. Since the noise realisations of the ABM and SPDE are independent, we cannot directly compare these individual simulations. Instead, we will compare the clustering dynamics of these models using a total of $10^3$ Monte-Carlo simulations. We found that further increasing the number of simulations had a negligible impact on our results.

Next, we compare the dynamics of the ABM, the SPDE, and also of the SPDE with linearised noise: in this last object (which is a popular choice due to its simpler implementation) the noise terms appearing in \eqref{spde_ReducedNoFeedSPDE} swap $\rho$ for $\bar{\rho}$, i.e., which is the solution to the following PDE
\begin{equation*}
    \partial_t \bar{\rho} = \frac{\sigma_s^2}{2} \Delta \bar{\rho} + \nabla \cdot [\bar{\rho} (b *\bar{\rho} + \nabla V)] .
\end{equation*}
Accordingly, the `linearised' SPDE takes the form
\begin{equation*}
    \partial_t \rho^L = \frac{\sigma_s^2}{2} \Delta \rho^L + \nabla \cdot [\rho^L (b *\rho^L + \nabla V)] + N^{-\frac{1}{2}} \sigma_s \nabla \cdot (\sqrt{\bar{\rho}} \, \xi_\rho) .
\end{equation*}
This alternative description has the advantage that the numerical implementation is simpler due to the additive noise.

\begin{figure}[tbh]
    \centering
    \includegraphics[width=0.99\linewidth]{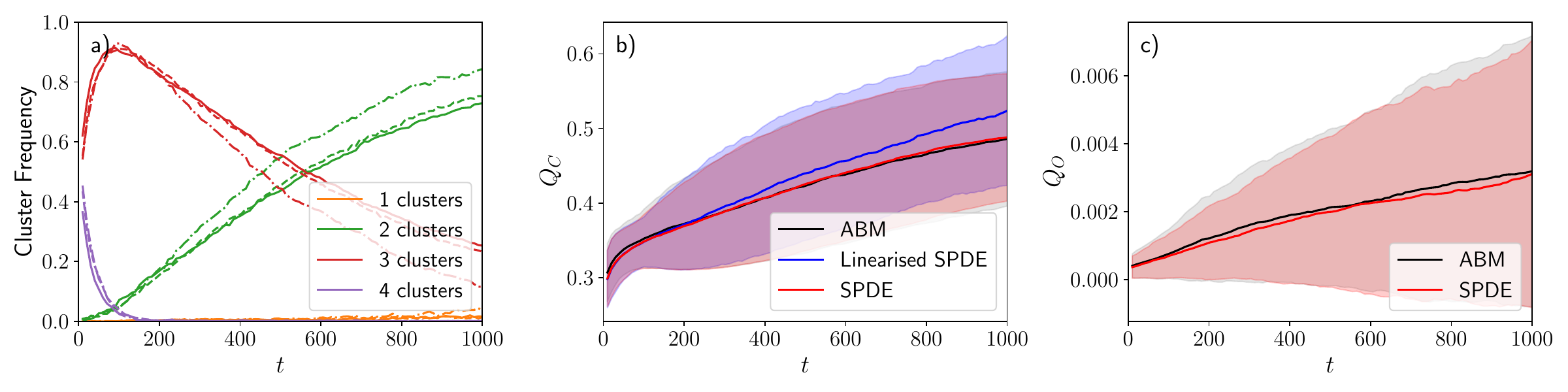}
    \caption{Cluster dynamics for the non-feedback model with $d=1$: ABM vs SPDE. The results are obtained using $N = 1000$ agents on the state space $\mathbb{T} \times \mathbb{R}$, with $1000$ independent realisations. a) The frequency of clusters for the ABM \eqref{particles_no_feedback} (solid line), reduced \eqref{spde_ReducedNoFeedSPDE} (dashed line) and linearised SPDE (dot-dashed line). b) The mean (solid line) and standard deviation (shaded area) of the clustering order parameter $Q_c$. c) The mean (solid-line) and standard deviation (shaded-area) of the opinion parameter $Q_o$}
    \label{fig:1DclusterStats}
\end{figure}

In Fig.~\ref{fig:1DclusterStats} we show the cluster dynamics of ABM, SPDE, and linearised SPDE, as characterized by the three quantities introduced in the previous section, that are: the relative frequency of clusters is presented in panel a), with the solid/dashed/dot-dash lines corresponding the the ABM/SPDE/linearised SDPE. The mean and standard deviation of $Q_C$ and $Q_o$ are shown in the panels b) and c), respectively. From these results we see that there is a close agreement between the ABM and SPDE for all three measures. Furthermore, we see that at long times the linearised model performs noticeably worse than the both ABM and SPDE.

\begin{figure}[tbh]
    \centering
    \includegraphics[width=0.99\linewidth]{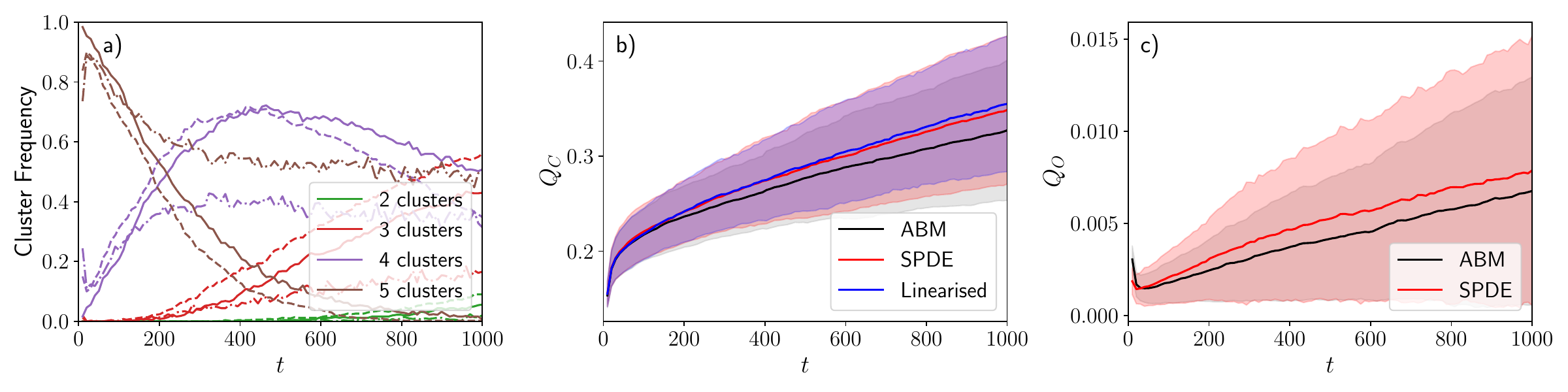}
    \caption{Cluster dynamics for the non-feedback model with $d=2$: ABM vs SPDE. The results are obtained using $N = 1000$ agents on the state space $\mathbb{T}^2 \times \mathbb{R}$, with $1000$ independent realisations. a) The frequency of clusters for the ABM \eqref{particles_no_feedback} (solid line), reduced \eqref{spde_ReducedNoFeedSPDE} (dashed line) and linearised SPDE (dot-dashed line). b) The mean (solid line) and standard deviation (shaded area) of the clustering order parameter $Q_c$. c) The mean (solid-line) and standard deviation (shaded-area) of the opinion parameter $Q_o$}
    \label{fig:2DclusterStatsBeta10}
\end{figure}

We extend this study to a two-dimensional social space in Fig.~\ref{fig:2DclusterStatsBeta10} for $R=0.15$. Once again, we see a close agreement between the clustering behaviour of the microscopic and mesoscopic models.  

\begin{figure}[tbh]
    \centering
    \includegraphics[width=0.99\linewidth]{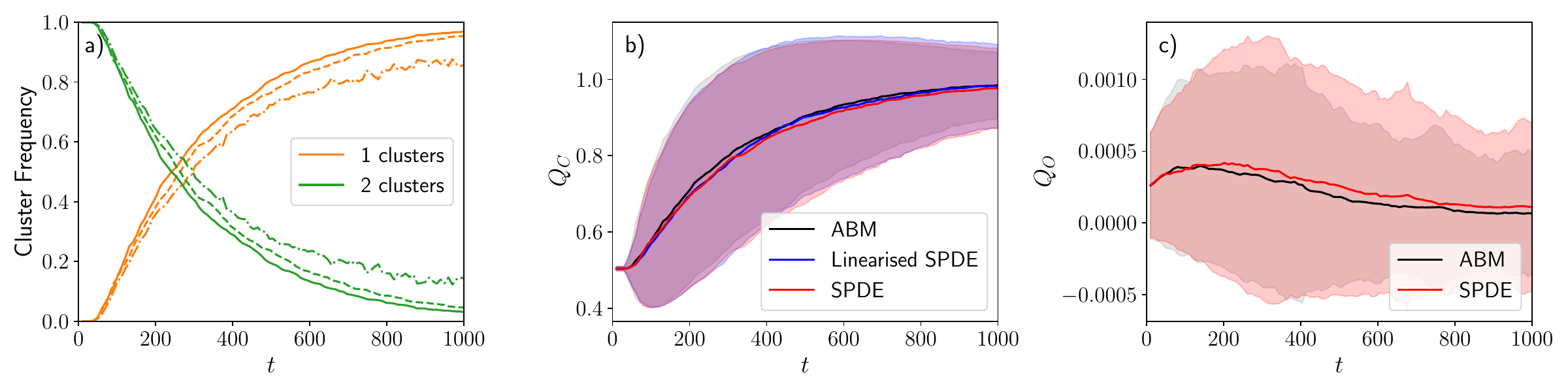}
    \caption{Cluster dynamics for the non-feedback model with an external potential and $d=1$: ABM vs SPDE. The external potential is given by the double-well \eqref{eq:ExtPotential_DWell}. The results are obtained using $N = 1000$ agents on the state space $\mathbb{T} \times \mathbb{R}$, with $1000$ independent realisations. a) The frequency of clusters for the ABM \eqref{particles_no_feedback} (solid line), reduced \eqref{spde_ReducedNoFeedSPDE} (dashed line) and linearised SPDE (dot-dashed line). b) The mean (solid line) and standard deviation (shaded area) of the clustering order parameter $Q_c$. c) The mean (solid-line) and standard deviation (shaded-area) of the opinion parameter $Q_o$}
    \label{fig:1DclusterStatsPot}
\end{figure}

Lastly, we consider the case when there is an external potential $V$ present. The form of the external potential is a double-well, centred at $x=0.5$,
\begin{equation}
    \label{eq:ExtPotential_DWell}
    V(x) = (s(x-0.5)^2 -h)^2.
\end{equation}
We set $s = 1$ and $h= 10^{-2}$, so that the potential has two minima at $x=0.4$ and $x = 0.6$ and the height of the barrier separating them is small enough for agents to move between the wells and form a single cluster in the time interval we consider. The dynamics of the clusters of the two systems are presented in Fig.~\ref{fig:1DclusterStatsPot}. Comparing these results with those of Fig.~\ref{fig:1DclusterStats}, we see that, due to the potential, only two clusters are formed initially. These two clusters then merge, and the merging rate computed from the SPDE is in good agreement with that observed in the ABM.

\subsection{Feedback model}

Let us now study the reduced \eqref{spde_ReducedFeedbackSPDE} for the feedback model. Unlike the previous case where a closing approximation only had to be made for one of the noise terms of the SPDE, for this model a closing approximation also had to be made for some of the drift terms. As such, we expect the reduced model to only be accurate when the closing approximation is valid, i.e. when all agents which are close in the social space have the same sign of opinion (see the discussion in Section \ref{subsec:FeedbackSPDE}). Throughout this section two sets of initial data are considered:
\begin{itemize}
    \item IC 1: In the social space agents are grouped into four distinct clusters of equal mass. The agents in a cluster share the same opinion and two of the clusters have opinion $\theta = 0.1$ while the remaining two have opinion $\theta = -0.1$. 
    \item IC 2: In the social space the agents are sampled from a uniform distribution on the interval $x \in \mathbb{T}$,  while the opinions are uniformly sampled from $\theta \in [-1,1]$.
\end{itemize}
For the first of these configurations, IC 1, the initial closing approximation \eqref{eq:ClErr} is minimised, while for the second, IC 2, it is maximised.

\begin{figure}[tbh]
    \centering
    \includegraphics[width=0.75\linewidth]{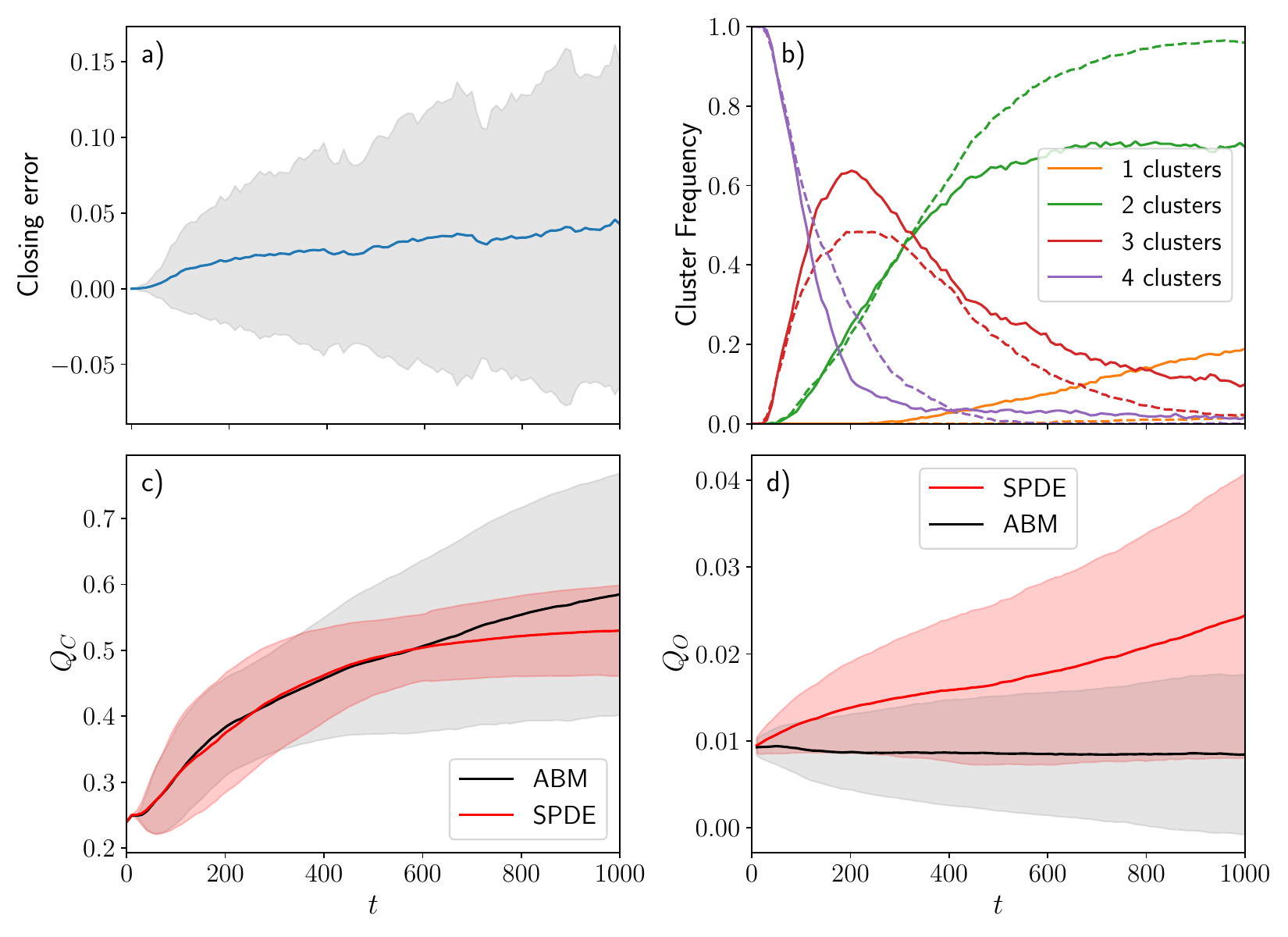}
    \caption{Cluster dynamics for the feedback model with initial condition IC 1: ABM vs SPDE. The results are obtained using $N = 1000$ agents on the state space $\mathbb{T} \times \mathbb{R}$, with $1000$ independent realisations. a) The mean (solid line) and standard deviation (shaded area) of the closing error \eqref{eq:ClErr} for the ABM \eqref{particles_feedback}. b) The frequency of clusters for the ABM (solid line) and reduced \eqref{spde_ReducedFeedbackSPDE} (dashed line) . c) The mean (solid line) and standard deviation (shaded area) of the clustering order parameter $Q_c$. d) The mean (solid-line) and standard deviation (shaded-area) of the opinion parameter $Q_o$}
    \label{fig:CoEsign_stats_4cl}
\end{figure}

In Fig.~\ref{fig:CoEsign_stats_4cl}, we compute the evolution of the closing error \eqref{eq:ClErr}, as well as the clustering behaviour, for IC 1. We see that $C_E$ has an initial value of zero which increases as the system evolves. This increase is due to the clusters merging, as agents with differing opinions overlap in the social space when this occurs. Comparing the clustering, we see that the SPDE is accurate in reproducing the ABM for early times, but gets worse for longer time-scales. This is in agreement with what we see in the time evolution for $C_E$.

\begin{figure}[tbh]
    \centering
    \includegraphics[width=0.75\linewidth]{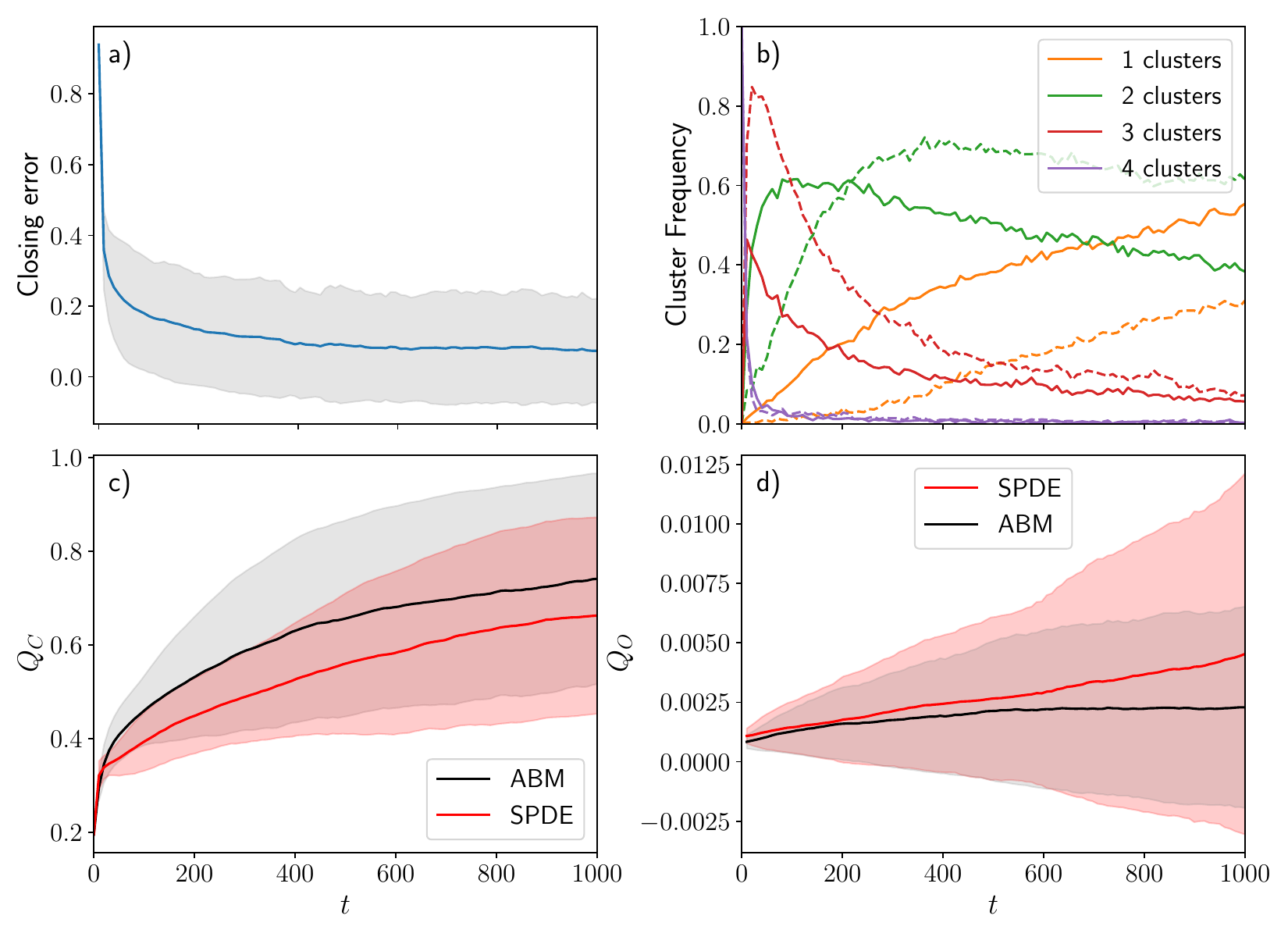}
    \caption{Cluster dynamics for the feedback model with initial condition IC 2: ABM vs SPDE. The results are obtained using $N = 1000$ agents on the state space $\mathbb{T} \times \mathbb{R}$, with $1000$ independent realisations. a) The mean (solid line) and standard deviation (shaded area) of the closing error \eqref{eq:ClErr} for the ABM \eqref{particles_feedback}. b) The frequency of clusters for the ABM (solid line) and reduced \eqref{spde_ReducedFeedbackSPDE} (dashed line) . c) The mean (solid line) and standard deviation (shaded area) of the clustering order parameter $Q_c$. d) The mean (solid-line) and standard deviation (shaded-area) of the opinion parameter $Q_o$}
    \label{fig:CoEsign_stats_Uni}
\end{figure}

Let us now perform a similar experiment for the case when $C_E$ is initially maximised, i.e. IC 2. In Fig.~\ref{fig:CoEsign_stats_Uni}a) we show the time-evolution of the closing error. We can see that from an initial value of one, it decreases rapidly. This is due to the cluster formation in the social space and the fact that a local consensus is reached within the clusters. The clustering behaviour of the ABM and SPDE is compared in the remaining panels of the figure. We see a much worse agreement between ABM and SPDE than in the previous figure. This is however expected as we choose the initial condition specifically such that the closing approximation made to derive the SPDE does not hold. Given that the closing error shows such a rapid decrease from the initially uniform data, it is sensible to allow the dynamics of the ABM to evolve for a short period of time before using the SPDE to characterise the long-term behaviour.

\begin{figure}[tbh]
    \centering
    \includegraphics[width=0.75\linewidth]{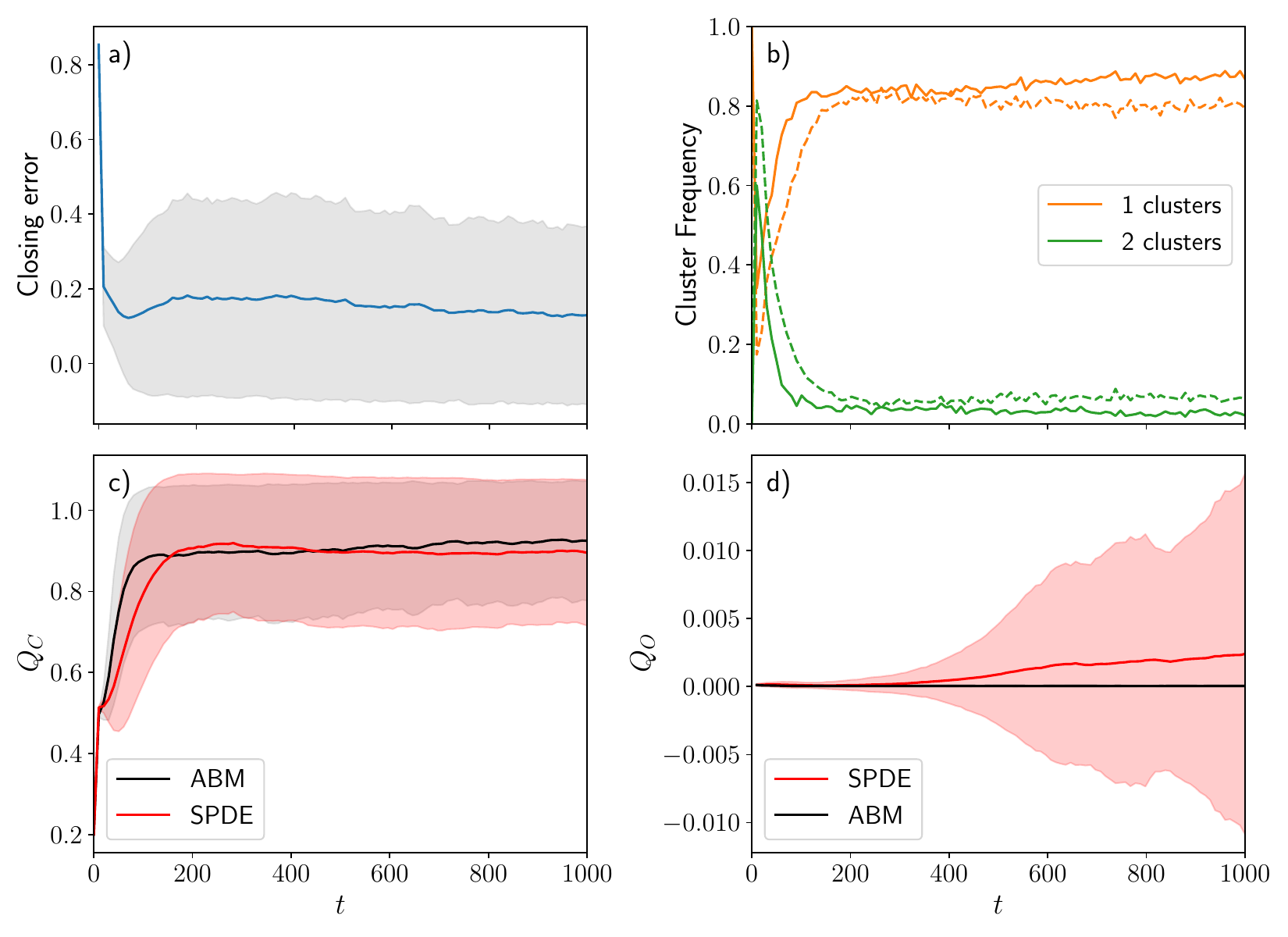}
    \caption{Cluster dynamics for the feedback model with external potential and initial condition IC 2: ABM vs SPDE. The external potential is given by the double-well \eqref{eq:ExtPotential_DWell}. The results are obtained using $N = 1000$ agents on the state space $\mathbb{T} \times \mathbb{R}$, with $1000$ independent realisations. a) The mean (solid line) and standard deviation (shaded area) of the closing error \eqref{eq:ClErr} for the ABM \eqref{particles_feedback}. b) The frequency of clusters for the ABM (solid line) and reduced \eqref{spde_ReducedFeedbackSPDE} (dashed line) . c) The mean (solid line) and standard deviation (shaded area) of the clustering order parameter $Q_c$. d) The mean (solid-line) and standard deviation (shaded-area) of the opinion parameter $Q_o$}
    \label{fig:CoEsign_stats_Pot}
\end{figure}

As the final experiment in this section we consider the case when there is an external potential influencing the dynamics. We will use the double-well potential \eqref{eq:ExtPotential_DWell}, as in Fig.~\ref{fig:1DclusterStatsPot}, together with initial data IC 2 from Fig.~\ref{fig:CoEsign_stats_Uni}. The closing error as well as cluster behaviour are presented in Fig.~\ref{fig:CoEsign_stats_Pot}. Due to the potential, two clusters form around the two minima of the potential, resulting in a rapid decrease of $C_E$. From the plot of the cluster frequency for the ABM, panel b), we see that these two clusters will then merge to form a single cluster. The results for the SPDE show fairly good agreement with the ABM, especially for the order parameter $Q_c$. The cluster frequency of the SPDE at longer times is, however, incorrect: it predicts that approximately $80 \%$ of the simulations form a single cluster, whereas in the ABM a single cluster forms approximately in $90 \%$ of the simulations. This discrepancy is also reflected in the opinion difference, since for the ABM a single cluster means that there is no difference. Nonetheless, broadly speaking, the SPDE shows similar behaviour to the ABM, performing better than in the case without external potential (compare the results of Fig.~\ref{fig:CoEsign_stats_Uni}).    

\subsection{Computational cost}

Let us briefly discuss how the reduced SPDE models \eqref{spde_ReducedNoFeedSPDE} and \eqref{spde_ReducedFeedbackSPDE} compare to alternative continuum descriptions of the ABM. 

\begin{figure}[tbh]
    \centering
    \includegraphics[width=0.9\linewidth]{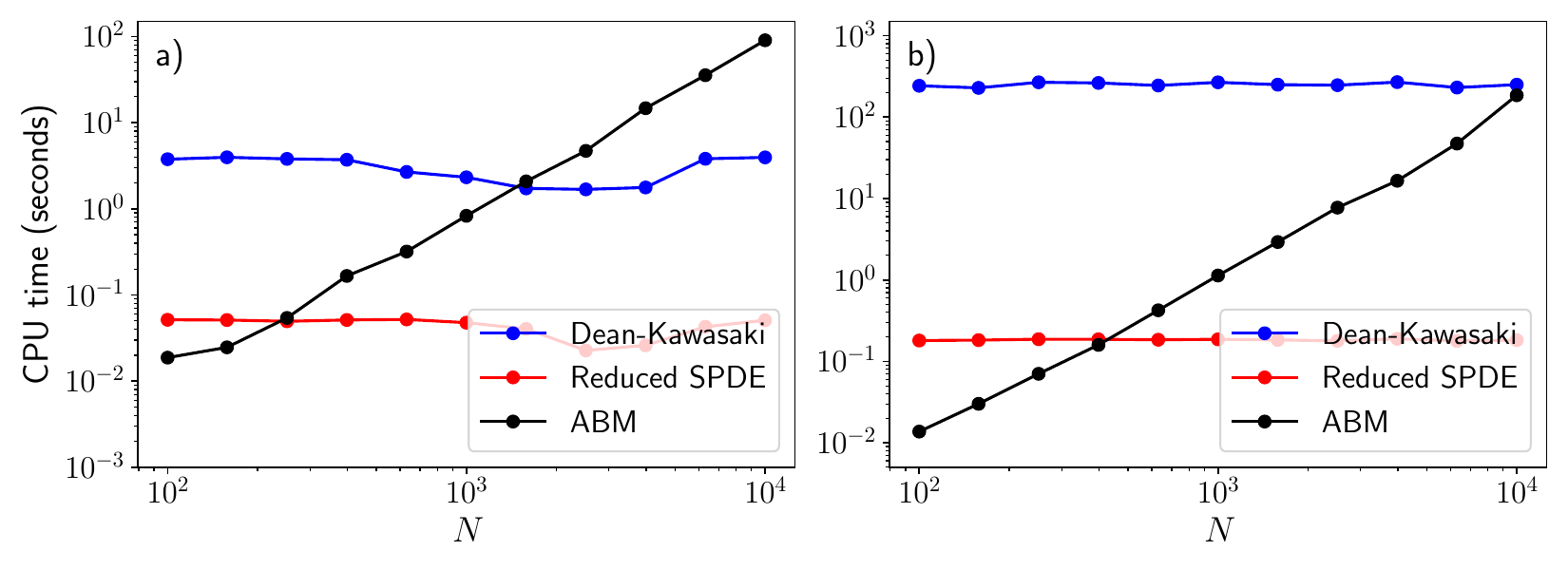}
    \caption{Comparison of the computational cost across models. The results show the simulation time (in seconds) required by the ABM, the {Dean--Kawasaki} equation, and the reduced SPDE to reach the final time $T = 10$, as the number of agents $N$ varies. Results for the non-feedback and feedback models are shown in panels a) and b), respectively}
    \label{fig:Time_Diff}
\end{figure}

Another alternative description is the full Dean--Kawasaki equation for the ABMs \cite[Eqn.~(25)]{djurdjevac2022feedback} 
\begin{equation}
    \label{eq:CoE_DK}
    \begin{aligned}
        \partial_t f(x,\theta,t) &= \frac{\sigma_s^2}{2} \Delta_x f(x,\theta,t)  + \nabla_x \cdot [f(x,\theta,t)(B(x,\theta,f_t) + \nabla_x V)]  \\
        &+\frac{\sigma_o^2}{2} \Delta_\theta f(x,\theta,t) +\nabla_\theta [f(x,\theta,t) (A(x,\theta,f_t) )] \\
        &+ N^{-\frac{1}{2}} \sigma_s \nabla_x \cdot (\sqrt{f(x,\theta,t)} \, \xi_\rho) + N^{-\frac{1}{2}} \sigma_o \nabla_\theta  (\sqrt{f(x,\theta,t)} \, \xi_j) ,
    \end{aligned}
\end{equation}
where 
\begin{equation*}
    \begin{aligned}
        B(x,\theta,f_t) &: = \int_{\mathbb{T}^d\times \mathbb{R}} b(x-y,\theta,\eta) f_t(dy,d\eta) \\
        A(x,\theta,f_t) &: = \int_{\mathbb{T}^d\times \mathbb{R}} a(x-y) (\eta - \theta) f_t(dy,d\eta) .
    \end{aligned}
\end{equation*}
This is an SPDE in $d+1$-dimensions describing the evolution of the full empirical measure \eqref{eq:FullEmpMeasure}. In contrast to the reduced feedback model, it does not require any closure approximation. We therefore expect this full description to reproduce the dynamics more accurately than the reduced models. This benefit, however, comes at the cost of a higher computational complexity. To illustrate this, we present in Fig.~\ref{fig:Time_Diff} the CPU time (in seconds) to compute a single simulation of the ABM, reduced SPDE and Dean--Kawasaki equation over the time-interval $t \in [0,10]$ for a varying number of agents $N$. The results without and with feedback are shown in panels a) and b), respectively. Due to the pairwise interactions the time required for the ABM scales super-linearly with $N$, while both continuum descriptions are independent of it. We see that the reduced SPDE is approximately $100$-times faster than the full description justifying its use when long-term dynamics for large $N$ is investigated.

\subsection{Real-world data}

In this section, we apply our approach to investigate the dynamics of opinion and social clusters in a real-word system. We consider empirical data from the General Social Survey (GSS) \cite{davern2024gss}, a large-scale survey collection of popular beliefs, attitudes and behaviours of the USA population since 1972. Recent work \cite{djurdjevac2024co} has shown that the feedback ABM \eqref{particles_feedback} accurately reproduces the emergent opinion and social patterns observed in the GSS data. Following these results, we represent the social space using a two-dimensional political compass constructed from data on political party affiliation (ranging from Democrat to Republican) and political ideology (ranging from liberal to conservative). Opinions related to governmental engagement and political decision-making provide a setting in which a feedback ABM offers a natural microscopic description of the system, as has been demonstrated empirically in \cite{djurdjevac2024co}.  Following this setup, agents' opinions are taken from the ``HelpSick" data-set on the survey question ``Should the government cover medical bills?". The political compass data are rescaled to the domain $[-0.25,0.25]\times[-0.25,0.25]$ and the opinions to the interval $[-1,1]$. Building on the validity of the feedback ABM for this data-set, we study here to what extent our reduced SPDE model captures the same cluster dynamics. 

\begin{figure}[tbh]
    \centering
    \includegraphics[width=0.99\linewidth]{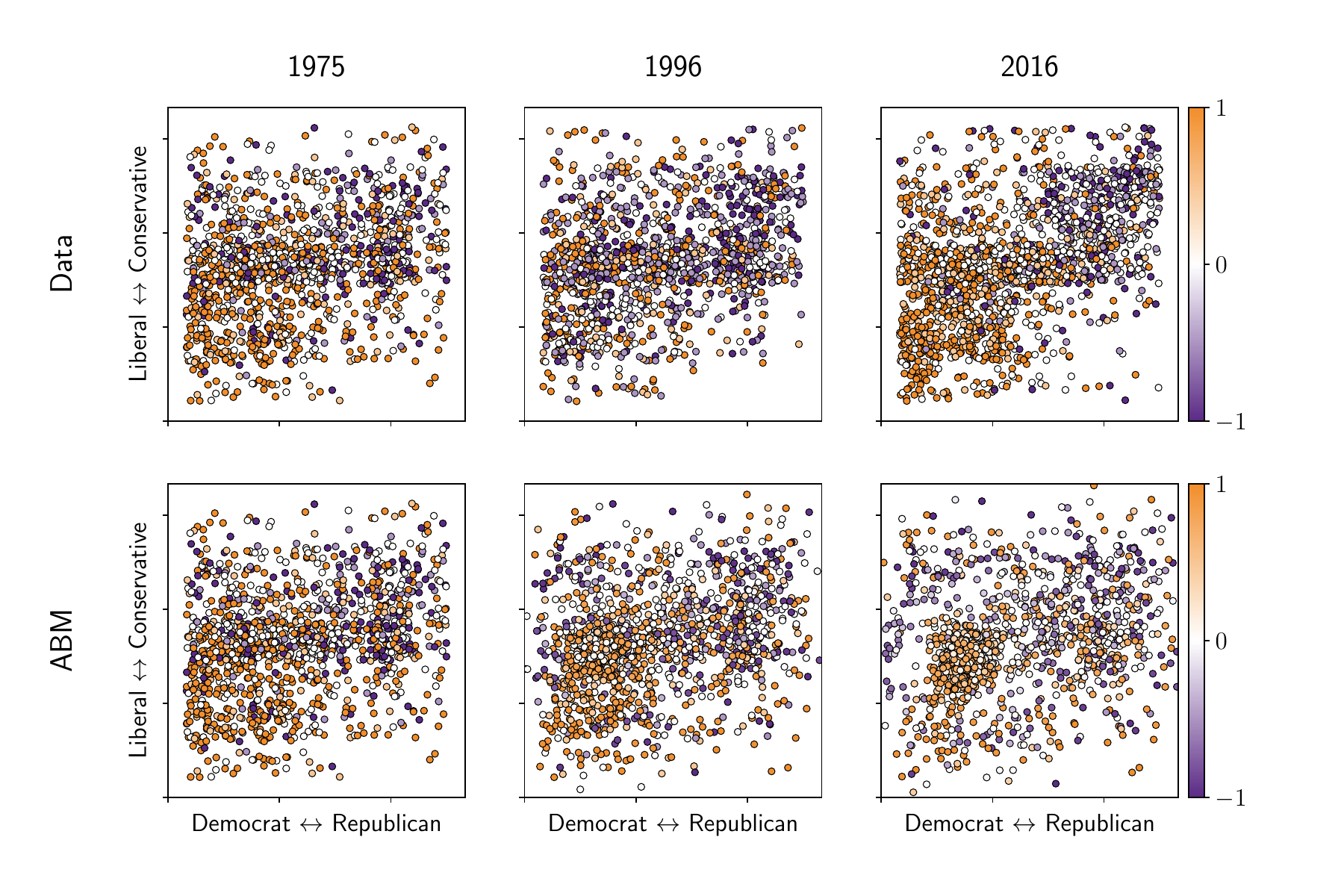}
    \caption{Comparison of the GSS data (top-row) and ABM simulation (bottom-row) for the ``HelpSick" dataset. The snapshots are for the years $1975$  (first column), $1996$  (second column) and $2016$ (third column). The social space is two-dimensional with the first dimension corresponding to party affiliation (from Democrat to Republican) and the second to political ideology (from liberal to conservative). The opinion of agents are coloured according to their response to the question ``Should the government cover medical bills?" using the colour-bars on the right of the figure}
    \label{fig:GSSdata_ABM}
\end{figure}

The number of agents is fixed to $N=1355$ the number of valid responses obtained in $1975$, which is the first year that the data is available for. The choice of parameters in the ABM is as in \cite{djurdjevac2024co}, i.e. $R_{o}=R_{s}=0.15$ and $\sigma_{o}=\sigma_{s}=0.05$. In Fig.~\ref{fig:GSSdata_ABM} we plot the snapshots of the GSS data and one ABM simulation for the initial year $1975$ and the presidential election years $1996$ and $2016$ in the United States. Towards $2016$, we observe a formation of one spatial and opinion cluster, both in the data and the simulation. Note that in the ABM simulation the number of agents is fixed, where in the GSS data we observe large fluctuations in the number of valid answers, see \cite{djurdjevac2024co} for more details. 

\begin{figure}[tbh]
    \centering
    \includegraphics[width=0.9\linewidth]{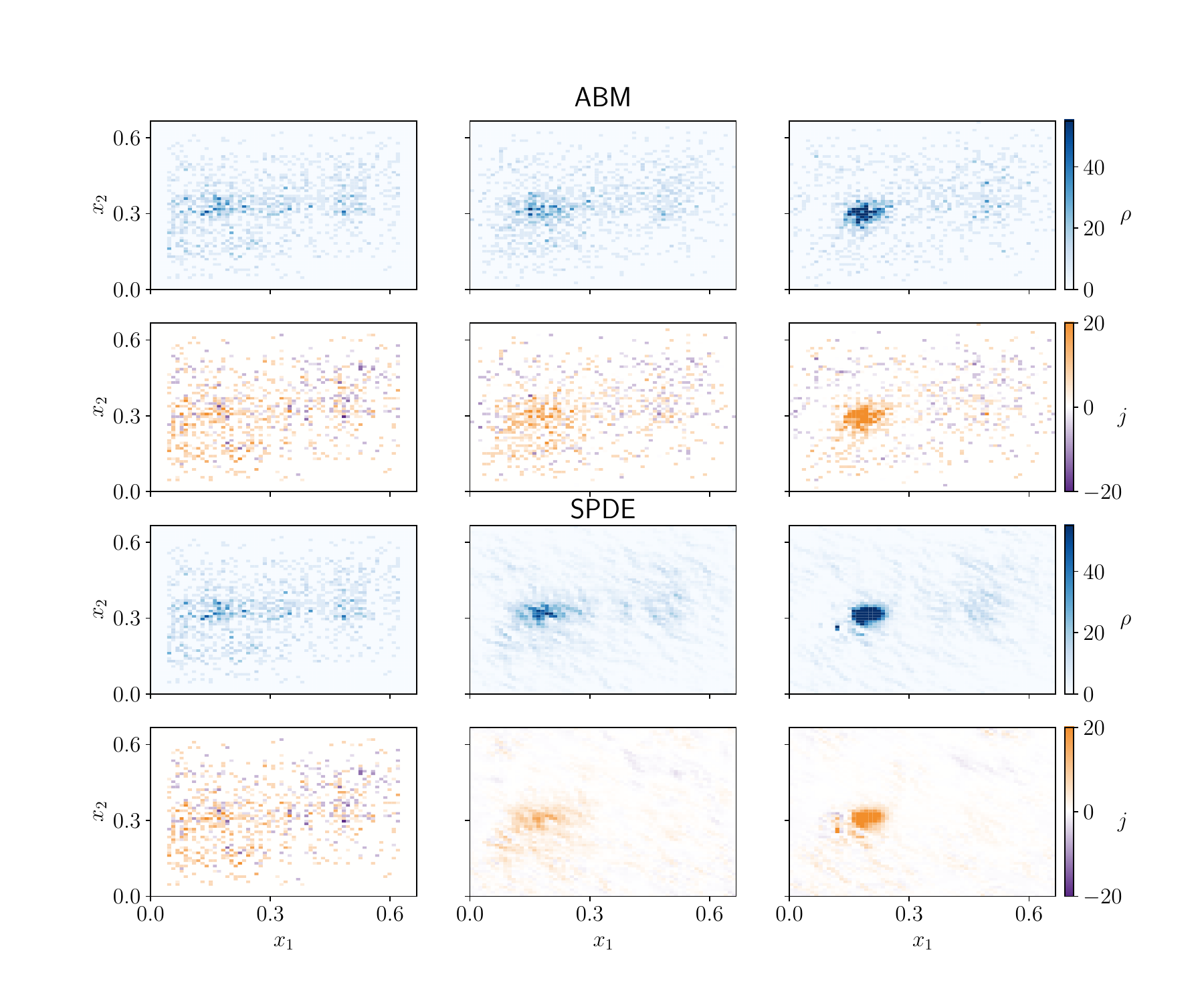}
    \caption{Comparison of the ABM and SPDE using the parameters for fitting the GSS data. Three snapshots are shown for the years $1975$ (first column), $1996$ (second column) and $2016$ (third column). The results for the ABM are shown in the top two rows, while those for the SPDE are in the bottom two rows}
    \label{fig:GSSdata_rho_j}
\end{figure}

In Fig.~\ref{fig:GSSdata_rho_j} we present snapshots of the empirical distributions for both the ABM and SPDE under the same parameter settings. We implment no-flux boundary conditions for the social space. The smoothed empirical densities of the ABM are displayed in the top two rows, while the bottom two rows show the solution of the SPDE, $\rho$ and $j$. We see that the SPDE successfully reproduces the clustering behaviour observed in the ABM, including the formation of a single cluster with a positive opinion (indicated by the {orange} shading of $j$). This example demonstrates that our SPDE formulation provides an accurate approximation of the underlying feedback-driven microscopic system and captures the emergent cluster dynamics well. Moreover, the SPDE model offers a computationally efficient framework, which makes it well suited for the analysis of large-scale real-world datasets, where fully resolved agent-based simulations become computationally expensive.

\section{Conclusion}
\label{sec:Conclusion}

In this work, we formally derived and studied reduced SPDE models that reproduce the clustering behaviour of ABMs at a significantly lower computational cost. The ABMs under consideration describe the co-evolution of agents' opinions and their positions in a social space, enabling the study of how clusters in the social space  influence opinion formation, and vice-versa. A central feature of these models is the dynamics of clusters in the social space and the evolution of the opinions within them.

We examined two settings: one in which the opinions are coupled to the social dynamics, but the social dynamics evolve independently of the opinions, and another in which a feedback mechanism couples the two. In both cases, the reduced models are obtained by tracking the evolution of the empirical density and the opinion-weighted empirical density. The former captures the spatial distribution of agents, while the latter describes the distribution of opinions among nearby agents in the social space.

The structure of the underlying ABMs determines the type of closure approximations required to obtain the SPDE models. In the absence of feedback, only the noise terms need to be closed, which we achieve using a hierarchy of densities. Although this approach does not fully capture the correlation structure of the noise, the reduced model shows good agreement with the ABM in our numerical experiments. Moreover, the form of the hierarchy is compatible with higher-order moment closures (cf.~\cite{Cornalba2023}), suggesting a natural direction for future work, namely incorporating correlation information and analysing higher-order statistics. When feedback is present, both drift and noise terms require closure. Here, we employ a mono-kinetic ansatz to obtain the SPDE description. This approach introduces ratios of densities, which can lead to numerical instabilities in regions of low density. Developing stable numerical schemes for such systems therefore remains an important open problem. Nonetheless, our numerical results indicate a good agreement between the ABM and SPDE. 

Overall, the proposed SPDE models successfully capture the key clustering behaviour of the underlying ABMs, while being significantly more computationally efficient than both the original ABMs and the associated Dean--Kawasaki equation. Although our focus has been on opinion dynamics, the methodology is sufficiently general to be applied to a broader class of interacting particle systems. Natural extensions of this work involve moving to more general settings, where interactions between agents include higher-order effects rather than being restricted to pairwise interactions, the noise is multiplicative rather than additive, and the domain is extended beyond the torus to account for boundary effects. Such models provide a more realistic description of social dynamics, but also introduce additional mathematical complexity, which we leave for future investigation. \\

{\bfseries Funding.} This work has been partially funded by the Deutsche Forschungsgemeinschaft (DFG, German Research Foundation) under Germany's Excellence Strategy -- The Berlin Mathematics Research Center MATH+ (EXC-2046/1, EXC-2046/2, project ID: 390685689).
\\

{\bfseries Data Availability Statement.} All source codes are available on request.  \\

{\bfseries Competing Interests.} The authors declare no competing interests.

\bibliographystyle{abbrv}
\bibliography{refs}

\section{Appendix}

\subsection{Derivation of reduced models for the non-feedback case}
\label{sec:App_NoFeedback}

The derivation of the reduced SPDE for the non-feedback model \eqref{spde_ReducedNoFeedSPDE} with one dimensional opinion variable is formal and proceeds as follows. For notational simplicity, we omit the explicit time dependence of the variables.

An application of It\^o's formula to the empirical density and the general divergence formula $\nabla \cdot (a\mathbf{b}) = \nabla a \cdot \mathbf{b} + a \nabla\cdot \mathbf{b}$ gives 
\begin{equation*}
    \begin{aligned}
        d \rho = &  N^{-1}\sum_{i=1}^{N}{\nabla \delta(x-X^i)} \cdot \left( N^{-1}\sum_{j=1}^N b(X^i-X^j)dt +   \nabla V(X^i) dt - \sigma_s dW_{t}^i \right) \\
        &+ \frac{1}{2} \sigma_s^2 N^{-1}\sum_{i=1}^{N}{\Delta \delta(x-X^i)} dt \\
        & =: \sum_{i=1}^4 b_i .
    \end{aligned}
\end{equation*}
It is straightforward to see that 
\begin{equation*}
    \begin{aligned}
        b_1 + b_2 =&  \frac{1}{N} \sum_{i=1}^N  \nabla  \delta(x-X^i) \cdot (b* \rho + \nabla V)(X^i) dt =  \int \nabla \delta(x-y) \cdot (b* \rho + \nabla V)(y) \rho(y) dy dt \\
        =&  \nabla \cdot [\rho (b *\rho + \nabla V)] \, dt ,        
    \end{aligned}
\end{equation*}
and
\begin{equation*}
    b_4 = \frac{1}{2} \sigma_s^2  N^{-1}\sum_{i=1}^{N}{\Delta \delta(x-X^i)} dt = \frac{1}{2} \sigma_s^2  \Delta \rho \,dt .
\end{equation*}
Introducing the vector noise field
\begin{equation*}
    Z_\rho(x,t) :=\frac{1}{N} \sum_{i=1}^N \delta (x - X_t^i) dW_t^i ,
\end{equation*}
we can write the noise term $b_3$ in divergence form, i.e.
\begin{equation*}
    b_3(x,t) = -\sigma_s \nabla \cdot Z_\rho(x,t) . 
\end{equation*}
The covariance of the noise field is formally given as
\begin{equation*}
    \mathbb{E}[Z_{\rho,\alpha}(x,t)Z_{\rho,\beta}(y,s)] = \frac{1}{N} \rho(x,t)
    \delta_{\alpha,\beta}\delta(x-y)\delta(t-s) ,
\end{equation*}
where the subscripts $\alpha$ and $\beta$ indicate the components of the vector. Since $Z_\rho$ is Gaussian we can consider the noise term that is equal in law
\begin{equation*}
    Z_\rho \overset{L}{=} N^{-\frac{1}{2}} \sqrt{\rho} \, \xi_\rho,
\end{equation*}
where $\xi_\rho$ is a vector space-time white noise. Combining all these terms gives us the SPDE for $\rho$. For $j$ another application of It\^o's formula gives
\begin{equation*}
    \begin{aligned}
        d j = &  -N^{-1}\sum_{i=1}^{N} N^{-1} \sum_{j=1}^N a(X^i-X^j) (\theta^j-\theta^i) \delta(x-X^i) dt   +  N^{-1}\sum_{i=1}^{N} \delta(x-X^i) \sigma_o dB_t^i \\
        & +  N^{-1}\sum_{i=1}^{N} \theta^i \nabla \delta(x-X^i) \cdot \left( N^{-1} \sum_{j=1}^N b(X^i-X^j) dt + \nabla V(X^i) dt \right)  \\
        &- N^{-1}\sum_{i=1}^{N} \theta^i \nabla \delta(x-X^i) \sigma_s \cdot dW_t^i  + N^{-1}\sum_{i=1}^{N} \theta^i \frac{1}{2}  \sigma_s^2 \Delta \delta(x-X^i) =: \sum_{i=1}^6 c_i .
    \end{aligned}
\end{equation*}
From the properties of the convolution (see discussion in \cite{zimper2025reduced}), we obtain
\begin{equation*}
    c_1 = - \rho (a *j) \, dt + j (a * \rho) \, dt,
\end{equation*}
and 
\begin{equation}\label{approx_c3}
    \begin{aligned}
        c_3 + c_4 = &   \frac{1}{N} \sum_{i=1}^N \theta^i  \nabla \delta(x-X^i) \cdot (b* \rho + \nabla V)(X^i) dt \\
        = &  \int \nabla \delta(x-y) \cdot (b* \rho + \nabla V)(y) j(y) dy dt \\
        =&  \nabla \cdot [j (b *\rho + \nabla V)] \, dt.
    \end{aligned}
\end{equation}
A direct computation shows that
\begin{equation*}
    c_6 =  \frac{1}{2} \sigma_s^2 \Delta j \, dt .
\end{equation*}
Introducing the vector random field
\begin{equation*}
    Z_j(x,t) :=\frac{1}{N}\sum_{i=1}^N \theta_t^i \delta (x-X_t^i) dW_t^i ,
\end{equation*}
and scalar noise field
\begin{equation*}
    Y_j(x,t) := \frac{1}{N} \sum_{i=1}^N \delta (x-X_t^i) dB_t^i ,
\end{equation*}
we can rewrite the noise terms as 
\begin{equation*}
    c_2(x,t) + c_5(x,t) = \sigma_o Y_j(x,t) -\sigma_s \nabla \cdot Z_j(x,t) , 
\end{equation*}
noting that the second noise term is naturally of divergence form, while the first is not. Their formal covariances are
\begin{equation*}
    \begin{aligned}
            \mathbb{E}[Y_j(x,t)Y_j(y,s)] & = \frac{1}{N} \rho(x,t) \delta(x-y) \delta(t-s) ,\\
    \mathbb{E}[Z_{j,\alpha}(x,t)Z_{j,\beta}(y,s)] & = \frac{1}{N}K(x,t) \delta_{\alpha,\beta}\delta(x-y)\delta(t-s) 
    ,
    \end{aligned}
\end{equation*}
where
\begin{equation*}
    K(x,t) := \frac{1}{N} \sum_{i=1}^N (\theta_t^i)^2 \delta(x-X_t^i) .
\end{equation*}
We can therefore write
\begin{equation*}
    Y_j \overset{L}{=} N^{-1/2} \sqrt{\rho} \,  {\xi_j} ,\qquad Z_j  \overset{L}{=} N^{-1/2} \sqrt{K}  \eta ,
\end{equation*}
where $\xi_j$ is a scalar white noise coming from the $B^i$ noise in $\theta^i$ and $\eta$ is space-time white noise, coming from the $W^i$ noise in $X^i$. Hence noises $\xi_\rho$ and $\eta$ come from the same particle Brownian motions $W^i$ and so are generally correlated: While it is clear that the space-time white noise $\xi_j$ should be independent of the noises $\xi_\rho$ and $\eta$, the correlation structure between $\xi_\rho$ and $\eta$ is not clear. In the absence of a precise characterization of this correlation, we adopt the simplifying closure $\eta=\xi_\rho$.  We emphasize that this identification is formal and is introduced only for convenience; it is not derived from the particle system and may not represent the correct correlation structure. Note that similar computations apply when deriving the SPDE in multi-dimensional opinion space, even though we do not treat this case explicitly.  

Having identified each of the terms we now have
\begin{equation}
    \label{eq:ReducedSPDE}
    \begin{aligned}
        \partial_t \rho &= \frac{\sigma_s^2}{2} \Delta \rho + \nabla \cdot [{ \rho (b *\rho + \nabla V) }] + N^{-\frac{1}{2}} \sigma_s \nabla \cdot (\sqrt{\rho} \, \xi_\rho)  , \\
        \partial_t j &= \frac{\sigma_s^2}{2} \Delta j - \rho (a *j) + j (a * \rho) + \nabla \cdot [ { j (b *\rho + \nabla V )}] + N^{-\frac{1}{2}} \sigma_o \sqrt{\rho} \, \xi_j + N^{-\frac{1}{2}}  \sigma_s \nabla \cdot (\sqrt{K} \, \xi_\rho) ,
    \end{aligned}
\end{equation}
where $K$ is not closed. Continuing with this hierarchy we obtain
\begin{equation*}
    \begin{aligned}
    dK = & \frac{1}{N} \sum_{i=1}^N 2 \theta^i \delta(x-X^i) d\theta^i + \delta(x-X^i) \sigma_o^2 dt - (\theta^i)^2 \nabla \delta(x-X^i) \cdot dX^i + (\theta^i)^2 \frac{1}{2} \sigma_s^2 \Delta \delta(x-X^i) dt  \\
     = & \frac{1}{N} \sum_{i=1}^N 2 \theta^i \delta(x-X^i) \left(-\frac{1}{N} \sum_{j=1}^N a(X^i-X^j) (\theta^j-\theta^i) dt + \sigma_o dB_t^i \right) + \delta(x-X^i) \sigma_o^2 dt \\
    &+ (\theta^i)^2 \nabla  \delta(x-X^i) \cdot \left(\frac{1}{N} \sum_{j=1}^N b(X^i-X^j)dt + \nabla V(X^i)dt- \sigma_s dW_t^i \right) + (\theta^i)^2 \frac{1}{2} \sigma_s^2 \Delta \delta(x-X^i) dt \\
    &=: \sum_{i=1}^7 T_i 
    \end{aligned}
\end{equation*}
It is straightforward to see that
\begin{equation*}
    \begin{aligned}
        T_3 &= \sigma_o^2 \rho \, dt \\
        T_7 &= \frac{\sigma_s^2}{2} \Delta K \, dt
    \end{aligned}
\end{equation*}
and
\begin{equation*}
    \begin{aligned}
        T_1 = & - \frac{2}{N} \sum_{i=1}^N  \theta^i \delta(x-X^i) \frac{1}{N} \sum_{j=1}^N a(X^i-X^j) \theta^j dt  
         + \frac{2}{N} \sum_{i=1}^N  (\theta^i)^2 \delta(x-X^i) \frac{1}{N} \sum_{j=1}^N a(X^i-X^j) dt \\
         = & -\frac{2}{N} \sum_{i=1}^N  \theta^i \delta(x-X^i) (a*j) dt + \frac{2}{N} \sum_{i=1}^N  (\theta^i)^2 \delta(x-X^i) (a * \rho) dt \\
         = & -2 j (a*j) dt + 2 K (a * \rho) dt .
    \end{aligned}
\end{equation*}
Furthermore,
\begin{equation*}
    \begin{aligned}
        T_4 + T_5 =  \frac{1}{N} \sum_{i=1}^N (\theta^i)^2 \nabla \delta(x-X^i) \cdot (b* \rho + \nabla V )(X^i) dt 
        =  \nabla \cdot [K (b *\rho + \nabla V)] \, dt .
    \end{aligned}
\end{equation*}
For the two noise terms, $T_2$ and $T_6$, by formally computing their covariances, we find that
\begin{equation*}
    \begin{aligned}
        T_2 &= \frac{1}{N} \sum_{i=1}^N 2 \theta^i \delta(x-X^i) \sigma_o dB_t^i \overset{L}{=} 2 N^{- \frac{1}{2}} \sigma_o \sqrt{K} \, \xi_K \\
        T_6 &= - (\theta^i)^2 \sigma_s \nabla \delta(x-X^i) \cdot dW_t^i \overset{L}{=} N^{-\frac{1}{2}} \sigma_s \nabla \cdot (\sqrt{L} \, \eta_K),
    \end{aligned}
\end{equation*}
where $\xi_K$ and $\eta_K$ are scalar and vector valued space-time white noises, respectively, and the  fourth-moment field is
\begin{equation*}
    L := \frac{1}{N} \sum_{i=1}^N (\theta^i)^4 \delta(x-X^i) .
\end{equation*}
Therefore,
\begin{equation*}
    \begin{aligned}
    \partial_t K &= \frac{\sigma_s^2}{2} \Delta K - 2 j (a*j) + 2 K (a * \rho) + \sigma_o^2 \rho  + \nabla \cdot [ { K (b *\rho + \nabla V )}] \\
    & + 2 \sigma_o N^{-\frac{1}{2}} \sqrt{K} \xi_K + N^{-\frac{1}{2}} \sigma_s \nabla \cdot (\sqrt{L} \eta_K) .
    \end{aligned}
\end{equation*}
To obtain a closed system, we drop the last two noise terms in the above expression:  our rationale is that, since these terms carry an additional pre-factor of $N^{-1/2}$ (which cumulates with the prefactor $N^{-1/2}$ appearing for the $K$-dependent noise term in \eqref{eq:ReducedSPDE}), they amount to higher-order contributions for the equations in \eqref{eq:ReducedSPDE}.

\begin{remark}\label{remark-multi-d-opinion}
    This derivation -- together with relevant noise equivalence procedures -- can be extended in a multi-dimensional opinion space (say, with $D$ opinion variables). This would require analysing a polynomial number $P(D)$ of $d$-dimensional SPDEs in order to keep track of all pairs of dependencies across the various components of the opinion space. This is far less costly than considering a full $d+D$-dimensional full model. We do not pursue this path in this paper.
\end{remark}

\subsection{Derivation of reduced models for the feedback case}
\label{sec:App_feedback}

Let us now derive the reduced model for the feedback model \eqref{particles_feedback}. As discussed in Section \ref{subsec:FeedbackSPDE}, the nonlinearity $\text{sgn}(\theta^i \theta^j)$ prevents certain drift terms from being expressed solely in terms of $\rho$ and $j$. To demonstrate how a reduced model can still be derived, we first formulate an unclosed system and then apply an appropriate closure approximation. 
From the computations in the previous subsection, it follows that only the terms 
\begin{equation*}
    \begin{aligned}
        b_1 &=   \frac{1}{N^2} \sum_{i=1}^N  \nabla \delta(x-X^i) \cdot\sum_{j=1}^N  b(X^i-X^j) \text{sgn}(\theta^i \theta^j) \, dt, \\
        c_3 &=  \frac{1}{N^2} \sum_{i=1}^N \theta^i \nabla \delta(x-X^i) \cdot\sum_{j=1}^N  b(X^i-X^j) \text{sgn}(\theta^i \theta^j) \, dt ,
    \end{aligned}
\end{equation*} 
are affected by the modified interaction. To obtain the unclosed system we introduce the auxiliary densities
\begin{equation*}
    m(x,t) := \frac{1}{N} \sum_{i=1}^N \text{sgn} (\theta_t^i) \delta(x-X_t^i) , \qquad q(x,t) := \frac{1}{N} \sum_{i=1}^N |\theta_t^i| \delta(x-X_t^i) ,
\end{equation*}
which allows us to rewrite, using the simple relations $\text{sgn}(ab) = \text{sgn}(a)\text{sgn}(b)$ and $a\text{sgn}(a) = |a|$,
\begin{equation*}
    b_1 =   \frac{1}{N} \sum_{i=1}^N \text{sgn}(\theta^i)  \nabla \delta(x-X^i) \cdot (b* m )(X^i) dt   =   \nabla \cdot (m (b *m)) dt ,
\end{equation*}
and
\begin{equation*}
    c_3 =  \frac{1}{N} \sum_{i=1}^N \theta^i\text{sgn}(\theta^i)  \nabla \delta(x-X^i) \cdot (b* m )(X^i) dt   =   \nabla \cdot (q (b *m)) dt .
\end{equation*}
This gives us the (formally) exact equations
\begin{equation*}
    \begin{aligned}
        \partial_t \rho =& \frac{\sigma_s^2}{2} \Delta \rho + \nabla \cdot \left[ m (b *m) + \rho \, \nabla V  \right] + N^{-\frac{1}{2}} \sigma_s \nabla \cdot (\sqrt{\rho} \xi_\rho) \\
        \partial_t j =& \frac{\sigma_s^2}{2} \Delta j - \rho (a *j) + j (a * \rho) + \nabla \cdot \left[ q (b *m) + j \, \nabla V    \right] \\
        &+ N^{-\frac{1}{2}} \sigma_o \sqrt{\rho} \xi_j + c_5 .
    \end{aligned}
\end{equation*}
Notice that we have not closed the noise terms $c_5$ by introducing a new quantity $K$. This is because the closing approximation we use for $m$ and $q$ can also be applied to $c_5$ to obtain a statistically equivalent noise for which we do not need to compute $K$. 

We now proceed to close the system. To do so, we employ the mono-kinetic ansatz \eqref{eq:MonoApprox}, which implies that, for each $x \in \mathbb{T}^d$, the opinions are concentrated at $\theta = u(x,t) = j(x,t)/\rho(x,t)$. This yields the approximation $\theta^i_t \approx u(X_t^i,t)$ which, combined with the non-negativity of $\rho$, gives
\begin{equation*}
    \begin{aligned}
        m(x,t) \approx \frac{1}{N} \sum_{i=1}^N \text{sgn}(u(X_t^i,t)) \delta(x-X_t^i) = \text{sgn}(j(x,t)) \frac{1}{N} \sum_{i=1}^N  \delta(x-X_t^i) = \text{sgn}(j(x,t)) \rho(x,t),
    \end{aligned}
\end{equation*}
and, similarly,
\begin{equation*}
    \begin{aligned}
        q(x,t) \approx \frac{1}{N} \sum_{i=1}^N |u(X_t^i,t)| \delta(x-X_t^i) = |u(x,t)| \rho(x,t) = |j(x,t)| = j(x,t) \text{sgn}(j(x,t)) .
    \end{aligned}
\end{equation*}
This allows us to close the drift terms as
\begin{equation*}
    \begin{aligned}
        \nabla \cdot \left[ m (b *m)  \right] & \approx  \nabla \cdot \left[ \rho \,\text{sgn}(j)  \,( b * ( \rho \, \text{sgn}(j) ) ) \right] , \\ 
        \nabla \cdot \left[ q (b *m)  \right] & \approx  \nabla \cdot \left[  j  \,\text{sgn}(j)  \,( b * ( \rho \, \text{sgn}(j) ) )  \right] .
    \end{aligned}
\end{equation*}
To be consistent we will forego the use of $K$ in the noise term $c_5$ and instead express it in terms of $u$ instead. In particular, by applying the mono-kinetic ansatz we find that
\begin{equation*}
    K(x,t)=\int \theta^2 f(x,\theta,t)\,d\theta \approx \rho(x,t)u(x,t)^2.
\end{equation*}
The second moment is clearly no longer an independent quantity: it is fully determined by $\rho$ and $u$. For that reason, $K$ does not need its own equation and the spatial noise in the $j$-equation $Z_j$ is written as
 \begin{equation*}
     N^{-1/2}\bigl(u\sqrt{\rho}\,\xi_\rho \bigr) .
 \end{equation*}
It is also straightforward to check that under this approximation the correlation between $Z_\rho$ and $Z_j$,
\begin{equation*}
        \mathbb{E} \left[ Z_{\rho,\alpha}(x,t)Z_{j,\beta}(y,s)  \right] = N^{-1}  \delta_{\alpha , \beta} \delta(t-s)  j(x,t) \delta(x-y)  ,
\end{equation*}
matches that between the corresponding noise terms of the SPDE, $N^{-1/2}\sqrt{\rho}\,\xi_\rho $ and $N^{-1/2} \, u\sqrt{\rho}\,\xi_\rho$.

\end{document}